\documentclass[a4paper,10pt]{article}
\pdfoutput=1 

\usepackage{jcappub} 

\usepackage[T1]{fontenc} 

\usepackage{enumitem,amssymb}
\usepackage{newtxtext,newtxmath}
\usepackage{lipsum,amsmath}
\usepackage{geometry}
\usepackage{kantlipsum,widetext}
\usepackage{rotating}
\usepackage{graphicx}
\usepackage{subcaption}
\usepackage{tabularx}
\usepackage{multirow}
\usepackage{booktabs}

\usepackage[normalem]{ulem}
\usepackage{xcolor}

\newcommand{\dd}{\mathrm{d}}
\newcommand{\cond}[2]{\left(#1\middle|#2\right)}
\newcommand{\pmean}{\bar{\mu}}

\title{Cosmology-dependent covariance in galaxy cluster number counts: consequences for parameter inference}

\author{
	Henrique C. N. Lettieri,$^{1}$\thanks{E-mail: henrique.cnl23@uel.br}
	Mariana Penna-Lima,$^{2,3}$\thanks{E-mail: pennalima@unb.br}
	Sandro D. P. Vitenti$^{1}$\thanks{E-mail: vitenti@uel.br}
	\\}

\affiliation[1]{Departamento de Física, Universidade Estadual de Londrina, Rod. Celso Garcia Cid, Km 380, 86057-970, Londrina, Brazil}
\affiliation[2]{Instituto de Física e Centro Internacional de Física, Universidade de Brasília, 70297-400, Brasília, Brazil}
\affiliation[3]{Programa de Pós-Graduação em Física, Universidade Estadual de Londrina, Rod. Celso Garcia Cid, Km 380, 86057-970, Londrina, Brazil}

\emailAdd{henrique.cnl23@uel.br}
\emailAdd{pennalima@unb.br}
\emailAdd{vitenti@uel.br}

\abstract{Galaxy clusters provide powerful constraints on cosmology through their
abundance as a function of mass and redshift. Parameter inference from cluster counts
requires modelling the data covariance entering the likelihood, including contributions
from Poisson shot noise and super-sample covariance (SSC) induced by long-wavelength
density fluctuations. While both the mean counts and their covariance depend on
cosmology, evaluating the full covariance during parameter inference can be
computationally expensive, particularly for SSC terms. As a result, many analyses adopt
approximations in which the covariance is computed at a single fiducial cosmology,
either from simulations or theoretical predictions, and kept fixed during inference. In
this work, we investigate the impact of constructing the covariance matrix at an
incorrect fiducial cosmology and quantify how this assumption propagates into the
estimation of $\Omega_c$, $\sigma_8$, and $w$. We perform a systematic analysis in
which the covariance is either varied consistently with the sampled cosmology or fixed
at displaced cosmological models, and we also explore intermediate strategies in which
selected components, such as the computationally expensive SSC term, are kept fixed
while others vary. Our analysis includes observational effects relevant for optical
surveys, including mass–proxy scatter and photometric redshift uncertainties, and
focuses on configurations representative of LSST-like surveys. We find that the
estimators of $\Omega_c$, $\sigma_8$, and $w$ remain unbiased even when the covariance
is constructed at an incorrect cosmology; however, fixing the covariance can lead to
significant over- or underestimation of credible regions. The magnitude and sign of
this effect are primarily driven by  parameters
controlling the amplitude of matter fluctuations, such as $\ln(10^{10}A_s)$ and $S_8$,
and for LSST-like surveys, an inconsistent covariance specification can artificially
modify the apparent $S_8$ tension inferred from cluster counts. We further show that a
single covariance update evaluated at the recovered best-fit cosmology is sufficient to
restore the correct uncertainties. These results indicate that while fixed-covariance
approximations may remain adequate for some single-probe analyses, a fully
cosmology-dependent treatment is required for consistent multi-probe studies.}

\arxivnumber{2606.30459}

\begin{document}

\maketitle

\section{Introduction}
\label{sec:introduction}

Galaxy cluster number counts are a powerful probe of cosmology, as the abundance of
massive halos and their spatial distribution are sensitive to both the growth of
structure and the expansion history of the Universe~\cite{Allen2011,Kravtsov2012}. A
key challenge, however, is that cluster mass is not directly observable. Cluster
samples are therefore selected through mass proxies such as X-ray observables, optical
richness, gravitational lensing, and Sunyaev-Zel'dovich (SZ) signals, whose connection
to the underlying halo mass must be modelled statistically through a mass-observable
relation. In millimetre and X-ray surveys, where cluster catalogs typically contain at
most a few hundred to a few thousand objects, cosmological analyses are often performed
using Poisson-based likelihoods~\cite{Planck2016,Lee2025}, sometimes in an effectively
unbinned
form~\cite{Mantz2010,PennaLima2014,Bocquet2019,Chiu2023,Ghirardini2024,Bocquet2024}.

Wide-field optical surveys, on the other hand, have entered a regime of much larger
cluster samples. Surveys such as the Sloan Digital Sky Survey (SDSS)~\cite{Rozo2010,
	Costanzi2019}, the Dark Energy Survey (DES)~\cite{DESY12020,DESY32025}, Javalambre
Physics Accelerating Universe Astrophysical Survey
(J-PAS)~\cite{Maturi2023,Doubrawa2024}, and the Stage IV Vera C. Rubin Observatory
Legacy Survey of Space and Time (LSST)~\cite{lsst2012} and
\textit{Euclid}~\cite{Laureijs2011} detect tens of thousands of clusters selected
primarily through optical richness, also focusing on mass calibration provided by weak
gravitational lensing measurements. In this high-count regime, analyses are commonly
performed assuming a Gaussian likelihood in bins of observed richness and photometric
redshift, both for computational convenience and to enable a consistent treatment of
correlated uncertainties. Similar considerations apply to next-generation X-ray surveys
such as SRG/eROSITA, which will deliver cluster catalogs of comparable statistical
power~\cite{Artis2025}.

Independently of whether one adopts a binned or an unbinned likelihood, the practical
implementation of cluster number count analyses remains computationally demanding. The
theoretical predictions require evaluating multi-dimensional integrals over the true
halo mass and redshift, as well as over observational quantities such as photometric
redshift estimates and mass proxies, each described by their own probability
distributions~\cite{Mantz2010, Bocquet2019, Costanzi2019}. In addition to modelling the
mean counts, a consistent likelihood must include an accurate description of the
covariance matrix.

The covariance of cluster counts is composed of a shot-noise (Poisson) contribution and
a SSC term \cite{hu2003sample, Valageas2011, Takada2013, takada2014joint,
	lacasa2016combining, Lacasa2019, Beauchamps2022}. The SSC arises from long-wavelength
density fluctuations larger than the survey footprint, which modulate the local growth
of structure and induce correlations across mass and redshift bins. Its incorporation
into cluster-count likelihoods has motivated significant theoretical developments,
including the formalism of Refs.~\cite{Valageas2011, lacasa2018super} and extensions
beyond purely Gaussian approximations \cite{payerne2023testing, payerne2024towards}.
For Stage IV surveys such as LSST, \textit{Euclid} and SRG/eROSITA
\cite{fumagalli2021euclid, payerne2024towards,hofmann2017erosita}, where statistical
errors are substantially reduced, an accurate treatment of SSC and of the full
covariance structure becomes essential.

Despite these advances, consistently incorporating cosmology-dependent covariance terms
remains challenging. The mean cluster counts depend on cosmology through the halo mass
function and comoving volume element, while the SSC term further depends on halo bias
and the large-scale matter power spectrum amplitude. As a result, both the mean signal
and its covariance are intrinsically cosmology-dependent. In unbinned likelihoods,
correlations across mass and redshift must be handled in a computationally efficient
manner, while in binned Gaussian approaches the covariance matrix must be evaluated and
inverted repeatedly during parameter inference. These challenges have motivated
approximate treatments and fast covariance estimators \cite{lacasa2018super,
	Lacasa2019, lacasa2023efficient, payerne2024towards}, as well as alternative
statistical frameworks such as simulation-based or likelihood-free inference approaches
\cite{Ishida2015, Zubeldia2025, Cerardi2025, Ntampaka2025, Payerne2026}.

In practice, computational efficiency is essential in multi-probe large-scale-structure
analyses. To fully exploit the available data and mitigate systematic uncertainties,
one of the primary goals of surveys such as the DES, LSST, and \textit{Euclid} is to
combine galaxy clustering, weak lensing, cluster clustering, number counts, and their
cross-correlations within a unified
framework~\cite{DESY12020,DESY32025,Prat2023,Fumagalli2024}.

The computation of the covariance matrix relies either on data-driven approaches or on
numerical simulations~\cite{Hirata2004, Hartlap2006, eifler2009dependence,
	Mandelbaum2013, Krause2017, Friedrich2018}, which are generally generated at a single
fiducial cosmology, or on (semi-)analytical calculations~\cite{Valageas2011,
	Takada2013}. The validity of this approximation must be reassessed as surveys enter a
precision regime in which systematic effects and modelling assumptions contribute at a
level comparable to (or even exceeding) the statistical
uncertainties~\cite{Hartlap2006, eifler2009dependence, Dodelson2013, Friedrich2018}.

Several studies have emphasized that super-sample covariance can significantly affect
parameter uncertainties \cite{takada2014joint, lacasa2018super}. In the context of weak
lensing and large-scale structure, the cosmology dependence of the covariance has been
shown to modify credible regions at a non-negligible
level~\cite{eifler2009dependence, Dodelson2013, Friedrich2018}.

However, important questions remain regarding the statistical impact of fixing the covariance matrix to an incorrect cosmology in cluster number count analyses. While previous studies have demonstrated that adopting an inconsistent covariance matrix can affect cosmological parameter constraints and have progressively incorporated more realistic observational ingredients, its impact has not yet been comprehensively assessed over a broad range of fiducial cosmological parameter choices, survey characteristics, and observational assumptions.

Cosmology-dependent covariance matrices have already been incorporated into cosmological analyses based on real cluster catalogues, including number count analyses accounting for mass--observable scaling relations and photometric-redshift uncertainties~\cite{Costanzi2019, DESY12020}.  Fumagalli et al.~\cite{fumagalli2021euclid} compared constraints on the matter density parameter $\Omega_m$ and on the amplitude of matter density fluctuations $\sigma_8$ obtained using  covariance  matrices derived from simulations and  analytical  calculations. In addition, since both covariance matrices were  computed for a fiducial cosmology, they  investigated the impact of adopting covariance matrices computed for two additional (incorrect) cosmologies. They showed that using an inconsistent covariance matrix can lead to either an underestimation or an overestimation of parameter uncertainties.  This analysis was restricted to idealized halo catalogs with true masses and redshifts.
More recently, Fumagalli et al.~\cite{Fumagalli2024} extended these studies to cluster clustering, while Fumagalli et al.~\cite{Fumagalli2026} considered the joint analysis of cluster number counts and cluster clustering, including mass--observable scaling relations and photometric-redshift uncertainties. Nevertheless, the statistical impact of covariance misspecification on cosmological parameter inference across different fiducial cosmological parameter choices and observational scenarios remains to be established.

The impact of covariance modelling is particularly relevant in light of the so-called
$\sigma_8$ or $S_8$ tension~\cite{Abdalla2022,Pantos2026}, which refers to the
discrepancy between the amplitude of matter fluctuations inferred from early-Universe
probes, such as the cosmic microwave background, and that obtained from late-time
large-scale-structure measurements
\cite{Planck2018,Asgari2021,Dalal2023,Sunayama2024,DESY32025}. The parameter $S_8
	\equiv \sigma_8 \sqrt{\Omega_m/0.3}$ is often used to characterize this tension, as it
approximately follows the degeneracy direction of low-redshift probes. Since cluster
abundance measurements are highly sensitive to both $\sigma_8$ and $\Omega_m$,
inaccuracies in the covariance modelling can artificially broaden or shrink credible
regions in the $\sigma_8$--$\Omega_m$ plane, potentially affecting the interpretation
of consistency (or tension) between different cosmological datasets (see, e.g.,
\cite{Miyatake2025} and references therein).

Motivated by these considerations, we adopt a forward-modelling approach to
systematically assess the impact of a cosmology-dependent covariance matrix on cluster
cosmological constraints. Using the \texttt{NumCosmo}~\cite{DiasPintoVitenti2014}
framework, whose C-based implementation enables the repeated evaluation of the cluster
pipeline and covariance modelling required for this study, we generate mock cluster
catalogs and compare analyses performed with fixed and varying covariance matrices.
This allows us to quantify potential biases in the estimators of the cold dark matter
density parameter $\Omega_c$, $\sigma_8$, and the dark energy equation-of-state
parameter $w$, as well as changes in their inferred uncertainties.

Our analysis is carried out in different observational scenarios, including idealized
catalogs with true masses and redshifts, as well as more realistic cases incorporating
mass-observable relations and photometric redshift uncertainties. This allows us to
evaluate the robustness of our conclusions under progressively more realistic
assumptions.

We further investigate how fixing the covariance matrix to an incorrect cosmology
influences the derived $\sigma_8$ and $S_8$ constraints and their interpretation in the
context of current tension discussions. Finally, we analyze the effect of recomputing
the covariance matrix at the best-fit parameters obtained from an initial analysis
with an incorrect covariance model.

This paper is organized as follows. In section~\ref{sec:cl_counts}, we introduce the
theoretical framework for cluster number counts, including the associated likelihood
and covariance matrix, and describe their dependence on the underlying cosmological
model. In section~\ref{sec:catalogs}, we present the construction of the mock catalogs
and the different observational scenarios considered in this analysis. The results are
discussed in section~\ref{sec:results}, and we summarize our conclusions in
section~\ref{sec:conclusions}.

\section{Cluster Counts}
\label{sec:cl_counts}

In this section, we present the theoretical framework for modelling cluster number
counts, i.e. the one-point statistics of the cluster distribution, over a survey region
of solid angle $\Omega_\text{sky}$, and describe its dependence on the underlying
cosmological parameters. We then account for the impact of SSC, which arises from
density fluctuations on scales larger than the survey volume and induces additional
correlations in the number counts. In the following subsections, we define the Gaussian
likelihood adopted in this work to connect the theoretical predictions to the mock
cluster catalogs, and describe how SSC is incorporated into the covariance of this
likelihood.

\subsection{Theoretical predictions}

The theoretical modelling of cluster number counts is based on the halo mass function,
which describes the comoving number density of dark matter halos per logarithmic mass
interval at a given redshift. It is defined as
\begin{equation}
	\label{eq:mass_function}
	\frac{\dd n(M,z)}{\dd\ln M}
	= \frac{\bar{\rho}_m}{M}\, \nu f(\nu)\,\frac{\dd \ln\nu}{\dd\ln M},
\end{equation}
where $\bar{\rho}_m$ is the mean comoving matter density of the Universe and $f(\nu)$
is the multiplicity function. In this work, we adopt the multiplicity function
calibrated by Tinker et al.~\cite{tinker2008toward}, which provides an accurate
description of halo abundances over a wide range of masses and
redshifts.
The peak height is defined as $\nu \equiv \delta_c / \sigma_R(z)$ with $\delta_c$
denoting the critical linear overdensity for collapse and $\sigma_R^2(z)$ the variance
of the linear matter density field smoothed with a top-hat filter of radius $R$
associated with the halo mass.

The value of $\delta_c$ depends on the assumed collapse dynamics and may, in general,
show a mild dependence on redshift and cosmology. An accurate interpolation capturing
this behaviour is provided in \cite{kitayama1996semi}. In this work, we adopt the
spherical top-hat collapse model, for which $\delta_c$ is approximately universal, and
we fix its value to $\delta_c = 1.686$.

The variance of the smoothed linear density field is computed as
\begin{equation}
	\label{eq:variance}
	\sigma_R^2(z) = \int_0^{\infty} \frac{\dd k}{2\pi^2}\,k^2\,P(k,z)\,|W(k, R)|^2,
\end{equation}
where $P(k,z)$ is the linear matter power spectrum extrapolated to redshift $z$, and
$W(k, R)$ is the Fourier transform of the window function. Throughout this analysis, we
employ a real-space top-hat filter, whose Fourier-space representation is
\begin{equation}
	\label{eq:window}
	W(k, R) = \frac{3}{kR}\,j_1(kR),
\end{equation}
with $j_1$ denoting the spherical Bessel function of first order.

The smoothing scale $R$ is related to the halo mass through a spherical overdensity
definition,
\begin{equation}
	\label{eq:mass_radius}
	M_\Delta = \frac{4\pi}{3}\,R_\Delta^3\, \Delta\rho_c,
\end{equation}
where $\rho_c$ is the critical density of the Universe. In this work, cluster masses
are defined with respect to an overdensity threshold of $\Delta = 200$.

The linear matter power spectrum is modelled as
\begin{equation}
	\label{eq:power_spectrum}
	P(k,z) = A_s\,k^{n_s}\,T^2(k)\,D^2(z),
\end{equation}
where $A_s$ is the amplitude of the primordial power spectrum, $n_s$ is the scalar
spectral index, $T(k)$ is the matter transfer function, and $D(z)$ is the linear growth
factor normalized to unity at the present epoch~\cite{PennaLima2014}. Throughout this
work, we adopt the Eisenstein-Hu transfer function~\cite{eisenstein1998baryonic}. This
approximation provides a fast evaluation of the linear matter power spectrum and is
sufficient for the purposes of this work, since both the mock catalogs and the
parameter inference are performed consistently using the same modelling. In analyses of
real observational data, a more accurate calculation of the linear power spectrum (e.g.
using Boltzmann solvers such as \texttt{CLASS} or \texttt{CAMB}) would be required.

While $A_s$ directly controls the normalization of the primordial power spectrum, it is
common in cluster abundance analyses to express this normalization in terms of the
root-mean-square fluctuation of the matter density field smoothed on a scale of
$8\,h^{-1}\mathrm{Mpc}$, denoted by $\sigma_8$. The two quantities are related by
\begin{equation}
	\label{eq:power_spectrum_normalization}
	A_s =
	\frac{\sigma_8^2}
	{\displaystyle
		\int_0^{\infty} \frac{\dd k}{2\pi^2}\,
		k^{n_s+2}\,T^2(k)\,W^2(k, 8\,h^{-1}\mathrm{Mpc})
	}.
\end{equation}

With the ingredients introduced above, the halo mass function in
eq.~\eqref{eq:mass_function} provides the theoretical prediction for the abundance
of galaxy clusters as a function of true mass and redshift. In practice, however,
cluster masses are not directly observable. Cluster surveys infer masses through one or
more observables (mass proxies), whose relation to the true halo mass is described
statistically by the conditional distribution $P(M_{\rm obs}\,|\,M,z)$.

Similarly, cluster redshifts can be measured either spectroscopically or
photometrically. While spectroscopic redshifts are highly precise, photometric
redshifts allow much larger samples at the cost of an additional scatter with respect
to the true redshift. This uncertainty is modelled through the photo-$z$ distribution
$P(z_{\rm ph}\,|\,z)$.

Taking these observational effects into account, the predicted distribution of clusters
in observed mass and redshift is obtained by convolving the halo mass function with the
mass-observable and photometric redshift relations:
\begin{equation}
	\label{eq:cluster_distribution}
	\begin{aligned}
		x(M_{\rm obs},z_{\rm ph})
		= \Omega_{\rm sky}\int \dd\ln M \int \dd z\;
		\frac{\dd^2V}{\dd z\,\dd\Omega}\,
		\frac{\dd n(M,z)}{\dd\ln M}\,
		P(M_{\rm obs}\,|\,M,z)\,
		P(z_{\rm ph}\,|\,z),
	\end{aligned}
\end{equation}
where $\dd^2V/(\dd z\,\dd\Omega)$ is the comoving volume element per unit solid angle.

\subsubsection*{Binned cluster number counts}

Rather than analysing the full continuous (unbinned) distribution, we perform a binned
analysis in observed mass and photometric redshift. The expected number of clusters in
a bin $[M_{{\rm obs},i},M_{{\rm obs},i+1}]$ and $[z_{{\rm ph},\alpha},z_{{\rm
							ph},\alpha+1}]$ is given by
\begin{equation}
	\label{eq:cluster_counts}
	\pmean_{i,\alpha}
	=\int_{M_{{\rm obs},i}}^{M_{{\rm obs},i+1}}
	\int_{z_{{\rm ph},\alpha}}^{z_{{\rm ph},\alpha+1}}
	x(M_{\rm obs},z_{\rm ph})\,
	\dd M_{\rm obs}\,\dd z_{\rm ph}.
\end{equation}

In order to isolate the impact of SSC, we consider three levels of realism in the
construction of mock catalogs. First, we analyze the idealized case of true masses and
redshifts, corresponding to delta-function relations
\begin{equation}
	P(M_{\rm obs}\,|\,M,z)=\delta(M_{\rm obs}-M),
	\qquad
	P(z_{\rm ph}\,|\,z)=\delta(z_{\rm ph}-z).
\end{equation}
We then progressively include mass-proxy scatter and photometric redshift
uncertainties.

\subsubsection*{Mass--observable relation}

To model the mass-proxy uncertainty, we adopt a log-normal relation between the true
halo mass and the cluster richness $\lambda$ (see, e.g., \cite{Murata2019, Costanzi2021}), described by
\begin{align}
	\label{eq:mass_richness_relation}
	P(\ln\lambda\,|\,M,z) & =
	\frac{1}{\sqrt{2\pi}\,\sigma(M,z)}
	\exp\!\left[
		-\frac{\bigl(\ln\lambda-\mu(M,z)\bigr)^2}{2\sigma^2(M,z)}
	\right],                                                                   \\[6pt]
	\label{eq:mass_richness_mean}
	\mu(M,z)              & = \mu_0+\mu_M\ln\!\left(\frac{M}{M_0}\right)
	+\mu_z\ln\!\left(\frac{1+z}{1+z_0}\right),                                 \\[6pt]
	\label{eq:mass_richness_variance}
	\sigma(M,z)           & = \sigma_0+\sigma_M\ln\!\left(\frac{M}{M_0}\right)
	+\sigma_z\ln\!\left(\frac{1+z}{1+z_0}\right).
\end{align}
Both the mass and the redshift dependences are expressed in logarithmic form. The former reflects the fact that the distribution of richness at fixed mass is approximately Gaussian in $\ln\lambda$, which makes $\ln M$ the natural conjugate variable; the latter closely follows the parameterizations adopted in the mass-richness relations of Refs.~\cite{Murata2019, Costanzi2021}. In these studies, the mean richness is parameterized as a polynomial in $\ln[(1+z)/(1+z_0)]$ around a pivot redshift $z_0$, including linear and quadratic terms, with a corresponding parameterization adopted for the scatter in Murata et al.~\cite{Murata2019}. Costanzi et al.~\cite{Costanzi2021} instead include an additional Poisson-like contribution in their model for the scatter. We also tested a linear parameterization in $z$ and found no significant difference in the results. We keep $\mu_z=\sigma_z=0$ throughout, so this choice affects only the form in which the redshift evolution would be parameterized, not the results presented here. In a full self-calibration approach, these nuisance parameters would be jointly constrained with cosmology. In this work, however, we keep them fixed to focus on the impact of SSC.

\subsubsection*{Photometric redshift uncertainties}

Photometric redshift measurements introduce an additional source of uncertainty in the
observed cluster distribution, as the inferred redshift $z_{\rm ph}$ differs from the
true redshift $z$ due to measurement scatter. We model this effect through a
conditional probability distribution $P(z_{\rm ph}\,|\,z)$, assumed to be Gaussian with
zero mean bias,
\begin{equation}
	\label{eq:photoz_distribution}
	P(z_{\rm ph}\,|\,z)=
	\frac{1}{\sqrt{2\pi}\,\sigma^{\rm ph}(z)}
	\exp\!\left[
		-\frac{(z_{\rm ph}-z)^2}{2\,\sigma^{\rm ph}(z)^2}
		\right].
\end{equation}
The photometric redshift scatter is taken to increase with redshift according to
\begin{equation}
	\label{eq:photoz_scatter}
	\sigma^{\rm ph}(z)=\sigma^{\rm ph}_0\,(1+z),
\end{equation}
where $\sigma^{\rm ph}_0$ sets the overall normalization of the photo-$z$ uncertainty.
In this work, we assume an unbiased photometric redshift relation and keep $\sigma^{\rm
		ph}_0$ fixed in all analyses.

Including this photo-$z$ probability distribution in
eq.~\eqref{eq:cluster_distribution} accounts for the migration of clusters across
redshift bins induced by photometric measurement scatter.

\subsubsection{Halo bias and its role in SSC}

Galaxy clusters are biased tracers of the underlying matter density field
\cite{kaiser1984spatial}. In particular, the abundance of halos within a finite survey
volume is sensitive to long-wavelength density fluctuations: a region that is slightly
overdense with respect to the cosmic mean effectively lowers the threshold for collapse
and enhances the formation of massive objects, leading to an excess of clusters
relative to the global expectation from the halo mass function alone. This modulation
is quantified by the halo bias function $b(M,z)$, which describes how halos of mass $M$
trace large-scale matter perturbations.

Since the halo bias can be constructed consistently within the same framework used to
model the halo mass function (e.g.~\cite{mo1996analytic}), we adopt the fitting
function calibrated in \cite{tinker2010large}, matching our choice of multiplicity
function. The halo bias plays a central role in the SSC contribution to the cluster
counts covariance, as it governs the response of the number counts to background
density modes larger than the survey window.

In direct analogy with the predicted binned abundance $\pmean_{i,\alpha}$, we define the
effective halo bias in each observed mass and photometric redshift bin $(i,\alpha)$ as
the abundance-weighted average of $b(M,z)$,
\begin{equation}
	\label{eq:bias_binned_direct}
	\begin{aligned}
		b_{i,\alpha}
		=
		\frac{\Omega_{\rm sky}}{\pmean_{i,\alpha}}
		\int_{M_{{\rm obs},i}}^{M_{{\rm obs},i+1}}
		\int_{z_{{\rm ph},\alpha}}^{z_{{\rm ph},\alpha+1}}
		\dd M_{\rm obs}\,\dd z_{\rm ph}
		 & \int \dd\ln M \int \dd z\;
		\frac{\dd^2V}{\dd z\,\dd\Omega}\,
		\frac{\dd n(M,z)}{\dd\ln M}
		\\[4pt]
		 & \times P(M_{\rm obs}|M,z)\,
		P(z_{\rm ph}|z)\,
		b(M,z)\; .
	\end{aligned}
\end{equation}
In practice, evaluating this quantity requires the same multidimensional convolutions
over mass and redshift that appear in the predicted cluster abundance $\pmean_{i,\alpha}$.
Because these integrals must be computed for each bin and for every sampled cosmology,
the effective bias constitutes one of the dominant computational costs of the pipeline.
It enters explicitly in the SSC contribution to the covariance matrix, where
super-survey density fluctuations couple coherently to the cluster number counts
through $b_{i,\alpha}$, inducing correlations across mass and redshift bins.
\subsection{Likelihood}
\label{subsec:likelihood}

The next step is to connect the theoretical predictions for the binned cluster number
counts, $\pmean_{i,\alpha}$, to the observed data (or mock catalogues) in order to constrain
the cosmological parameters.

The expression for the likelihood requires specifying the statistical model that links
the discrete observed counts to the underlying matter distribution. In the standard
approach, one assumes that the cluster counts arise from a Poisson sampling of a
continuous underlying density (or intensity) field. Given this field, the expected
number of objects in each bin, $\pmean_{i,\alpha}$, is fixed, and the counts in
different bins are statistically independent. In this sense, the Poisson model assumes
that all correlations between bins are fully encoded in the deterministic prediction
for $\pmean_{i,\alpha}$.

Under these assumptions, the probability of observing a discrete number of clusters in
each bin is described by a Poisson likelihood,
\begin{equation}
	\label{eq:poisson_likelihood}
	\mathcal{L}_{\rm P}\cond{N}{\pmean}
	=
	\prod_{i=1}^{B_M}
	\prod_{\alpha=1}^{B_z}
	\frac{
		\pmean_{i,\alpha}^{\,N_{i,\alpha}}
		e^{-\pmean_{i,\alpha}}
	}{
		N_{i,\alpha}!
	},
\end{equation}
where $B_M$ and $B_z$ denote the number of mass and redshift bins, respectively,
$N_{i,\alpha}$ is the observed number of clusters in bin $(i,\alpha)$, and
$\pmean_{i,\alpha}$ is the theoretical prediction given by
eq.~\eqref{eq:cluster_counts}.

In practice, however, the density field is itself
described statistically, and therefore the expected counts $\pmean_{i,\alpha}$ inherit
this stochasticity and should be treated as random variables with a joint distribution
$P(\pmean)$ induced by the underlying field.
The Poisson model then holds only conditionally, $P\cond{N}{\pmean}$, and the full
likelihood is obtained by marginalizing over $\pmean$,
\begin{equation}\label{PN}
	P(N)
	=
	\int \dd\pmean\;
	P(\pmean)\,
	\prod_{i,\alpha}
	\frac{
		\pmean_{i,\alpha}^{\,N_{i,\alpha}}
		e^{-\pmean_{i,\alpha}}
	}{
		N_{i,\alpha}!
	}.
\end{equation}
In this formulation, correlations between bins arise from the joint distribution
$P(\pmean)$, even though the counts are conditionally independent given $\pmean$. In
general, this marginal likelihood does not admit a simple closed form, which makes an
exact treatment of SSC challenging.

Approximate schemes have therefore been developed. In particular, \cite{Lacasa2019}
proposes a systematic expansion that captures the effect of these fluctuations on the
likelihood. In the following, we adopt a similar approach.

In the regime where the expected counts per bin are sufficiently large, the marginal
likelihood can be approximated by matching its first two moments (mean and covariance),
leading to a multivariate Gaussian description. We therefore adopt the Gaussian likelihood
\begin{equation}
	\label{eq:gaussian_likelihood}
	\mathcal{L}_{\rm G}(N)
	=
	\frac{1}{\sqrt{(2\pi)^{B}\det C}}
	\exp\!\left[
		-\frac{1}{2}
		\left(N-\langle\pmean\rangle\right)^{\rm T}
		C^{-1}
		\left(N-\langle\pmean\rangle\right)
		\right],
\end{equation}
where $B=B_M\times B_z$ is the total number of bins, $\langle\pmean\rangle = \langle N
	\rangle$ is the mean number of clusters under $P(\pmean)$, which follows directly
from eq.~\eqref{PN}, and $C$ is the covariance matrix of the binned counts, including
both Poisson and super-sample contributions. In this formulation, the impact of SSC
is entirely captured through additional contributions to the covariance matrix, as
discussed in the next subsection.

\subsection{Covariance Matrix}
\label{sec:covariance}

In the Gaussian approximation adopted in this work, all the information about
statistical fluctuations in the cluster number counts is encoded in the covariance
matrix. From eq.~\eqref{PN}, the covariance of the counts can be written exactly as~\footnote{For simplicity, hereafter we drop the comma between the mass and redshift indices, i.e., $\bar{\mu}_{i,\alpha} \equiv \bar{\mu}_{i\alpha}$.}
\begin{equation}
	\label{eq:total_covariance}
	C_{i\alpha,j\beta} \equiv \mathrm{Cov}(N_{i\alpha}, N_{j\beta}) = C^{\mathrm{SN}}_{i\alpha,j\beta} + C^{\mathrm{field}}_{i\alpha,j\beta},
\end{equation}
with
\begin{equation}
	\label{eq:shot_noise_covariance}
	C^{\mathrm{SN}}_{i\alpha,j\beta} = \delta_{ij}\delta_{\alpha\beta}\langle \bar{\mu}_{i\alpha} \rangle, \qquad C^{\rm field}_{i\alpha,j\beta} = \mathrm{Cov}(\pmean_{i\alpha}, \pmean_{j\beta}).
\end{equation}
\noindent The first term corresponds to Poisson shot noise, and the second term arises
from fluctuations of the underlying density field. Equation~\eqref{eq:total_covariance}
is an exact identity. It follows from applying the law of total covariance to the
hierarchical model of eq.~\eqref{PN}: given $\pmean$ the counts are Poisson and
conditionally independent, so $\mathrm{Cov}\cond{N_{i\alpha},N_{j\beta}}{\pmean} =
	\delta_{ij}\delta_{\alpha\beta}\,\pmean_{i\alpha}$, and averaging over $P(\pmean)$
yields eqs.~\eqref{eq:total_covariance}--\eqref{eq:shot_noise_covariance} with no
assumption on the functional form of $P(\pmean)$.

The approximation is made at the previous step, where the marginal likelihood
eq.~\eqref{PN} is replaced by the Gaussian form eq.~\eqref{eq:gaussian_likelihood},
i.e., where the likelihood is truncated at the first two moments of the counts. Once
that replacement is made,
eqs.~\eqref{eq:total_covariance}--\eqref{eq:shot_noise_covariance} provide all the
information required to construct the likelihood, and the problem reduces to computing
the covariance of the expected counts,
$C^{\rm field}_{i\alpha,j\beta}$.

In general, $C^{\rm field}_{i\alpha,j\beta}$ receives contributions from a broad range of scales and depends on the survey geometry and binning scheme. A full evaluation requires integrating the two-point statistics of the density field over the survey window.

In the following, we approximate this term by retaining only the contribution from
long-wavelength modes of the density field, leading to the SSC approximation. These
modes coherently modulate the expected counts across the survey volume, inducing
correlations between different bins. Under this approximation, the field covariance is
written as
\begin{equation}
	\label{eq:ssc_covariance}
	C^{\rm field}_{i\alpha,j\beta} \approx C^{\rm SSC}_{i\alpha,j\beta}
	=
	b_{i\alpha}\,\langle \pmean_{i\alpha} \rangle\;
	b_{j\beta}\,\langle \pmean_{j\beta} \rangle\;
	S_{\alpha\beta},
\end{equation}
where $b_{i\alpha}$ is the effective halo bias averaged over bin $(i,\alpha)$, and
$S_{\alpha\beta}$  is the covariance of the matter density contrast between redshift bins
$\alpha$ and $\beta$, averaged over the survey volume.

The restriction to long wavelengths is not a property of $S_{\alpha\beta}$ itself. As
shown below, $S_{\alpha\beta}$ is built from the full matter power spectrum weighted by
the survey window function, and therefore receives contributions from small-scale modes
as well; which modes actually contribute is set by the window. The SSC approximation
lies instead in eq.~\eqref{eq:ssc_covariance}, where the response of the counts to the
density field is modelled as a linear-bias response to a single density contrast taken
to be coherent across each bin. The volume averaging in
eq.~\eqref{eq:S_matrix_definition} then suppresses the contribution of modes that are
not coherent over the corresponding volume, that is, modes with wavelengths smaller
than the bin size, which is what makes the long-wavelength modes dominant in practice.

More generally, the cross-correlation of the matter density contrast at two redshifts can be written as
\begin{equation}
	\label{eq:density_covariance_general}
	\sigma^2(z_1,z_2)
	\equiv
	\langle \delta(z_1)\,\delta(z_2)\rangle
	=
	\int \frac{d^3k}{(2\pi)^3}\,
	P(k,z_1,z_2)\,
	W(\mathbf{k},z_1)\,
	W^\star(\mathbf{k},z_2),
\end{equation}
where $P(k,z_1,z_2)$ is the unequal-time matter power spectrum\footnote{The unequal-time power spectrum is computed as $P(k, z_1, z_2) = \sqrt{P(k, z_1) P(k, z_2)}$. From eq.~\eqref{eq:power_spectrum}, it can be computed replacing $D^2(z)$ by $D(z_1)D(z_2)$.} and $W(\mathbf{k},z)$ is the Fourier transform of the survey window function.

In general, the window function encodes the survey geometry and selection, including
the angular footprint, masking, and radial binning. Because $W$ couples angular and
radial modes, evaluating eq.~\eqref{eq:density_covariance_general} directly becomes
computationally demanding and strongly survey dependent. To address this, we follow the
approach developed in \cite{lacasa2018super}, where the covariance is expanded in
spherical harmonics. In this formalism, the clustering and geometrical contributions
factorize, and the dependence on the matter power spectrum is captured by the angular
matter power spectrum,
\begin{equation}
	\label{eq:angular_power_spectrum}
	C_\ell^m(z_1,z_2)
	=
	\frac{2}{\pi}
	\int dk\,k^2\,
	j_\ell(kr_1)\,
	j_\ell(kr_2)\,
	P(k,z_1,z_2),
\end{equation}
where $j_\ell$ denotes the spherical Bessel function and $r_i$ is the comoving distance
corresponding to redshift $z_i$.

The survey geometry enters separately through an angular window power spectrum
$C_\ell(W)$, such that the covariance can be written as
\begin{equation}
	\label{eq:partial_sky_covariance}
	\sigma^2(z_1,z_2)
	=
	\frac{1}{\Omega_{\rm sky}^2}
	\sum_{\ell}(2\ell+1)\,
	C_\ell(W)\,
	C_\ell^m(z_1,z_2),
\end{equation}
where $\Omega_{\rm sky}=4\pi f_{\rm sky}$ is the solid angle of the footprint, and
$C_\ell(W)$ is normalised so that $\sum_\ell(2\ell+1)C_\ell(W)=\Omega_{\rm sky}$ for a binary mask. This expression corresponds to the general partial-sky case, where masking and survey boundaries induce mode coupling through $C_\ell(W)$.

Finally, the matrix $S_{\alpha\beta}$ entering eq.~\eqref{eq:ssc_covariance} is
obtained by averaging this expression over the redshift bins of the likelihood,
\begin{equation}
	\label{eq:S_matrix_definition}
	S_{\alpha\beta}
	=
	\int \frac{dV_1}{V_\alpha}
	\int \frac{dV_2}{V_\beta}\,
	\sigma^2(z_1,z_2),
\end{equation}
where $V_\alpha$ denotes the comoving volume of redshift bin $\alpha$. In this step, we
isolate the contribution of modes that are coherent over the survey volume, effectively
averaging over smaller-scale fluctuations and retaining only the large-scale component
that enters the SSC approximation. This construction separates the dependence on the
large-scale density fluctuations, encoded in $S_{\alpha\beta}$, from the response of
the cluster abundance, described by the bias factors in eq.~\eqref{eq:ssc_covariance}.
In the analysis that follows, this separation allows us to efficiently explore the
impact of SSC on cosmological constraints by keeping $S_{\alpha\beta}$ fixed while
recomputing the bias-dependent contribution to the covariance, thereby avoiding the
need for a full evaluation of the field covariance at each likelihood evaluation.

Super-sample variance scales inversely with the volume enclosed by the survey, so to
leading order $S_{\alpha\beta} \propto 1/f_{\rm sky}$. We can test this expectation
directly, since the partial-sky expression makes no such assumption. We compute
$S_{\alpha\beta}$ from eq.~\eqref{eq:partial_sky_covariance}, adopting a circular
footprint (a spherical cap) whose solid angle matches the survey area of each layout,
similar to the treatment in appendix A of ref.~\cite{Costanzi2019}. The angular power
spectrum of the window, $C_\ell(W)$, is obtained from a \textsc{HEALPix} realization of
the cap at $N_{\rm side}=512$, and the sum over multipoles is truncated at
$\ell_{\max}=2N_{\rm side}$. Since $S_{\alpha\beta}$ depends on the mask only through
$C_\ell(W)$, the position of the cap on the sphere is irrelevant.

\begin{figure}[!ht]
	\centering
	\includegraphics{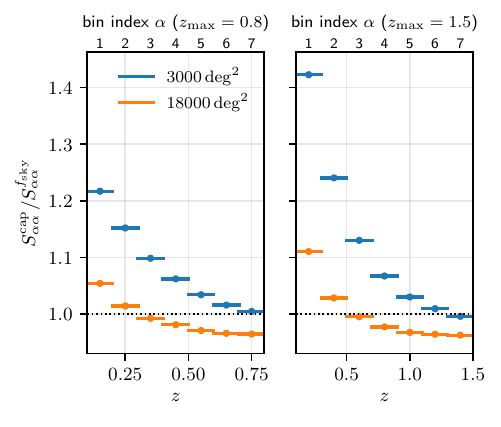}
	\caption{Diagonal elements of the SSC kernel computed on a spherical cap,
		$S^{\rm cap}_{\alpha\alpha}$, relative to the $1/f_{\rm sky}$ rescaling of the
		full-sky kernel, eq.~\eqref{eq:fullsky_sigma}, for $z_{\max}=0.8$ (left) and
		$z_{\max}=1.5$ (right). Each segment spans one redshift bin; the upper axis
		gives the bin index $\alpha$. The dotted line marks equality. The two treatments
		agree to within a few percent at high redshift and depart progressively towards lower redshift, where the footprint subtends a small comoving transverse scale.}
	\label{fig:footprint_S}
\end{figure}

The natural point of comparison is the full-sky kernel rescaled by $1/f_{\rm sky}$:
\begin{equation}
	\label{eq:fullsky_sigma}
	\sigma^2_{f_{\rm sky}}(z_1,z_2)
	=
	\frac{1}{2\pi^2\,f_{\rm sky}}
	\int dk\,k^2\,
	j_0(kr_1)\,
	j_0(kr_2)\,
	P(k,z_1,z_2),
\end{equation}
which reduces to the full-sky result when $f_{\rm sky}=1$. Figure~\ref{fig:footprint_S}
shows the ratio of the two kernels. They agree to a few per cent over most of the
redshift range. The agreement degrades systematically towards low redshift, reaching
$22\%$ for a configuration with $\Omega_{\text{sky}} = 3000\,\mathrm{deg}^2$ and maximum redshift $z_{\max}=0.8$ and $42\%$ for $z_{\max}=1.5$, in
the bins where the footprint subtends a small comoving transverse scale and the survey
volume becomes elongated along the line of sight. Thus, the $1/f_{\rm sky}$ scaling
captures the leading area dependence of the SSC kernel, but the angular structure of
the mask is not negligible in the lowest-redshift bins. We therefore evaluate the full
partial-sky expression throughout rather than adopting the simple $1/f_{\rm sky}$
rescaling.

A useful consequence follows from this scaling. The expected counts grow with the
footprint, $\langle\pmean_{i\alpha}\rangle \propto f_{\rm sky}$, while
$S_{\alpha\beta}\propto 1/f_{\rm sky}$ decreases with the survey area, so the
super-sample term of eq.~\eqref{eq:ssc_covariance} scales as $C^{\rm SSC}\propto f_{\rm
			sky}$, the same leading dependence as the shot-noise term $C^{\rm SN}\propto f_{\rm
			sky}$. The relative weight of the two contributions is therefore largely independent of
the survey area. Summed over the redshift bins of our fiducial configuration, the SSC
term contributes $2.26$ and $2.07$ times the shot-noise variance at $3000$ and
$18000\,\mathrm{deg}^2$ for $z_{\max}=0.8$, and $1.44$ and $1.24$ for
$z_{\max}=1.5$.%
\ Sensitivity to a
misspecified covariance is consequently set primarily by the depth and binning of the
survey rather than by its area. With this covariance model in place, we can now
investigate how SSC impacts cosmological constraints derived from cluster number
counts.

\section{Fiducial Models and Mock Cluster Catalogs}
\label{sec:catalogs}

To study the impact of fixing the covariance matrix to a specific cosmological model,
we generate mock cluster catalogs assuming a fiducial cosmology, from which cluster
redshifts and masses (or mass proxies) are drawn.  Because the input cosmology of the mock
catalogs is known exactly, this controlled setup allows us to quantify the biases
and uncertainties that arise when the covariance matrix is evaluated using an incorrect cosmological model.  We  can therefore directly compare the
results obtained when the covariance matrix is computed assuming the correct and incorrect cosmologies.

For the construction of the mock catalogs we follow the hierarchical resampling scheme
introduced in Refs.~\cite{PennaLima2014, Ishida2015} and implemented in
\texttt{NumCosmo}~\cite{DiasPintoVitenti2014}, to which we refer for the full algorithm.
For a given fiducial cosmology, we first draw the binned number counts $N_{i,\alpha}$ from
the multivariate Gaussian of eq.~\eqref{eq:gaussian_likelihood}, whose covariance comprises
both the Poisson shot-noise and the SSC contributions; the mock counts therefore carry the
same super-sample correlations between bins that the likelihood models. Conditional on these
counts, the individual cluster properties, mass (or mass proxy) and redshift, are then drawn
from the corresponding mass-observable and photometric redshift relations.

The mock catalogs are generated assuming fiducial cosmological models that specify the
true underlying properties of dark matter halos. Cluster number counts provide their
strongest constraints on the matter density parameter $\Omega_{m}$\footnote{In the
	figures, we denote the present-day value as $\Omega_{m0}$.} and the amplitude of matter
fluctuations $\sigma_8$, and, for surveys reaching $z \gtrsim 1.0$, also on the dark
energy equation-of-state parameter $w$. These parameters are therefore allowed to vary
over ranges commonly explored in Markov Chain Monte Carlo analyses.  Specifically, we assume uniform priors given by $\Omega_c \in [0.01, 0.9]$, $\ln(10^{10}A_s) \in [2.0, 5.0]$, and $w \in [-5, 0]$. This choice enables us to probe a broad region of cosmological parameter space and to test the robustness of our methodology under different cosmological scenarios.

All remaining cosmological parameters are kept fixed throughout the analyses, namely
the Hubble constant $H_0 = 67.81 \, \text{km/s/Mpc}$, $n_s = 0.9667$, the baryon density parameter $\Omega_b = 0.0486$, and massless neutrinos. These values are consistent with the Planck 2018 cosmological constraints~\cite{Planck2018}. The specific
combinations of cosmological parameters adopted in each fiducial scenario are presented
and discussed in sections~\ref{sec:s_matrix}, \ref{sec:mc}, \ref{sec:s8_tension} and \ref{sec:best_fit}.

For each fiducial cosmological model, we consider four different experimental setups by
varying two survey characteristics that strongly affect cluster number count analyses:
the sky coverage and the redshift depth. We adopt survey configurations motivated by
current and forthcoming optical surveys. Specifically, we consider an intermediate
survey area of $\Omega_\text{sky} = 3000 \, \text{deg}^2$ ($f_{\rm sky} = 0.073$),
characteristic of Stage-III experiments and partial-sky photometric observations, as
well as a wide survey covering $\Omega_\text{sky} = 18000 \, \text{deg}^2$ ($f_{\rm
			sky} = 0.436$), typical of Stage-IV experiments such as LSST~\cite{lsst2012}. For each
survey area, we explore two redshift ranges, $z \in [0.1, 0.8]$ and $z \in [0.1, 1.5]$,
resulting in four survey configurations per fiducial cosmological model.

In addition to the choice of fiducial cosmological parameters, survey area, and
redshift depth, we consider different levels of observational realism in the mock
catalogs. The first consists of halo catalogs with true masses and true redshifts,
corresponding to an idealized scenario in which these quantities are perfectly known.
The second includes mass proxies while retaining true redshifts, approximating the
level of information expected from spectroscopic surveys and from photometric surveys
such as J-PAS. The third, and most realistic, case incorporates both mass proxies and
photometric redshift estimates, as expected for wide-area photometric surveys such as
LSST and Euclid.

For the mass-observable relation, we use
eqs.~\eqref{eq:mass_richness_relation}--\eqref{eq:mass_richness_variance}
assuming: $\mu_0 = 3.207$, $\mu_M = 0.993$, $\mu_z = 0.0$ for the mean $\mu(M, z)$ and
$\sigma_0 = 0.456$, $\sigma_M = -0.169$ and $\sigma_z = 0.0$ for the variance
$\sigma(M, z)$ taken from \cite{murata2018constraints}. This choice is conservative because it corresponds to a larger intrinsic scatter than the values inferred in more recent analyses of Dark Energy Survey data~\cite{DESY32025}, thereby providing a more pessimistic characterization of the mass--observable relation while remaining
broadly consistent with expectations for richness-based mass estimates in photometric
surveys. Regarding the photo-z redshift uncertainties, we adopt $\sigma^{\text{ph}}_0 = 0.03$
in eq.~\eqref{eq:photoz_scatter}, as expected in surveys like LSST~\cite{mandelbaum2018lsst}.

In all cases, i.e., true or photometric redshift, the range in $z$ is divided into
seven bins equally spaced, chosen to ensure a sufficient number of clusters per bin and
the validity of the Gaussian likelihood approximation. Regarding the true mass, we use
a single bin spanning $M \in [10^{14} \, \text{M}_\odot, 10^{15} \, \text{M}_\odot]$.
Finally, for the mass-proxy, we consider four equally spaced bins in richness in the
range $\lambda \in [20, 200]$. 

The complete cluster pipeline is implemented within the \texttt{NumCosmo}
framework,\footnote{\url{https://github.com/NumCosmo/NumCosmo}} encompassing everything
from the background cosmology to the large-scale structure modelling, including the
halo mass function, halo bias, the mass-richness relation, photometric redshifts, and
cluster count predictions. Furthermore, the likelihood module incorporates a Gaussian
likelihood with Super-Sample Covariance (SSC) contributions, alongside a comprehensive
suite of statistical tools for Markov Chain Monte Carlo (MCMC) sampling, Monte Carlo
simulations, and best-fit optimization.

The $S$ matrix entering the covariance is computed with the super-sample covariance
module implemented in \texttt{NumCosmo}, which evaluates the angular power spectra of
eq.~\eqref{eq:angular_power_spectrum} with the non-Limber spherical-Bessel solver. All
matrices computed with \texttt{NumCosmo} were validated by direct integration and
reproduce the results of \texttt{PySSC}~\cite{lacasa2023efficient}. The
\texttt{NumCosmo} implementation and its validation will be described in detail in a
forthcoming publication (Vitenti et al., in preparation).

\section{Analyses and Results}
\label{sec:results}

In a realistic cluster analysis the cosmology is unknown, yet the covariance matrix
must be specified to evaluate the likelihood. Because the SSC term is expensive, the
covariance (or parts of it) is often built at a
single fiducial cosmology and held fixed while the cosmological parameters are
sampled. The question we address is how this approximation propagates into the
inferred parameters: does it bias the estimators, and does it distort the reported
uncertainties?

To answer this in a controlled way, we fix a single fiducial cosmology,
\begin{equation}
	\label{eq:fiducial_main}
	(\Omega_{m0},\, \sigma_8, \,w) = (0.3098,\,0.8159, \,-1.0),
\end{equation}
together with the remaining parameters of
section~\ref{sec:catalogs},\footnote{The sampled density parameter is the cold dark
	matter density $\Omega_c=0.2612$; throughout we quote the total matter density
	$\Omega_{m0}=\Omega_c+\Omega_b=0.3098$.} and generate
\emph{one} mock data vector per survey configuration. As described in
section~\ref{sec:catalogs}, we consider the
four survey layouts spanned by $\Omega_{\rm sky}\in\{3000,18000\}\,\mathrm{deg}^2$ and
$z_{\max}\in\{0.8,1.5\}$.

We study covariance misspecification in two complementary configurations. In the
\emph{forward} configuration the mock is held at the fiducial cosmology and the same data
vector is re-analysed under several constructions of the covariance matrix, displaced to
other cosmologies and labelled in section~\ref{sec:s_matrix}; because the data vector is
fixed, any change in the inferred posterior is attributable solely to the covariance
choice. In the \emph{mirror} configuration the converse holds: the
covariance is fixed at the fiducial value while the mock is generated at a displaced
cosmology, so that the data correspond to the unknown true one and the covariance to an
assumed cosmology.  If the impact of covariance misspecification depends only on the mismatch between the data and covariance cosmologies, rather than on which of the two is displaced, the two configurations should probe the same effect. The mirror configuration is therefore included as an internal consistency check of this interpretation, which we test explicitly in section~\ref{sec:mirror}. We use these labels, \emph{forward} and
\emph{mirror}, throughout.

We organise the study in two stages. We first map the effect on the posterior
\emph{width} using individual MCMC runs. In section~\ref{sec:s_matrix} we vary how the
covariance is built on a fixed (\emph{forward}) mock, isolating the role of the SSC $S$
matrix, of the full covariance, and of the shot-noise-only limit. In
section~\ref{sec:combined} we verify that shifting two cosmological parameters
simultaneously produces no effect beyond the sum of the individual shifts. In
section~\ref{sec:mirror} we show that the \emph{mirror} configuration reproduces
the \emph{forward} results once both are placed on a common cosmology-mismatch axis. We
then turn, in section~\ref{sec:mc}, to a Monte Carlo (MC) ensemble to show that covariance
misspecification does not add any detectable bias to the parameter \emph{estimators}, so
that its effect is confined to the parameter \emph{uncertainties}. In
section~\ref{sec:realism} we verify that these results are robust to a realistic mass proxy
and to LSST-like photometric redshifts. The consequences
for the $\sigma_8$/$S_8$ tension and a practical mitigation strategy are discussed in
sections~\ref{sec:s8_tension} and \ref{sec:best_fit}. Altogether, the systematic campaign
underlying sections~\ref{sec:s_matrix}--\ref{sec:realism} comprises $444$ complete
MCMC analyses ($474$ including sections~\ref{sec:s8_tension} and~\ref{sec:best_fit})
and more than $3\times10^4$ best-fit Monte Carlo realizations.

Throughout, we quantify the impact of a covariance choice on a parameter $\theta \in
	\{\Omega_{m0},\sigma_8,w\}$ through the ratio of its marginal posterior width to the
width obtained when the covariance is built consistently at the true (fiducial)
cosmology,
\begin{equation}
	\label{eq:width_ratio}
	R_\theta \;\equiv\; \frac{\sigma_\theta}{\sigma_\theta^{\rm match}},
	\qquad
	\Delta_\theta \;\equiv\; 100\,(R_\theta-1)\;[\%],
\end{equation}
where the reference analysis (\textsc{match}, defined in section~\ref{sec:s_matrix})
uses the covariance evaluated at the same cosmology that generated the mock. By
construction $R_\theta=1$ for the reference (\textsc{match});
$R_\theta<1$ ($>1$) signals an under- (over-)estimated uncertainty.

\subsection{Covariance choices on a fixed mock}
\label{sec:s_matrix}

For each survey layout we re-analyse the fiducial mock with the covariance $C$ of the
Gaussian likelihood (eq.~\eqref{eq:gaussian_likelihood}) built in one of three ways,
which we label and use throughout:
\begin{description}
	\item[\textsc{S-fix}] only the SSC kernel $S$ (eq.~\eqref{eq:S_matrix_definition})
	      is held fixed, while the bias and number-count prefactors
	      $b_{i\alpha}\bar\mu_{i\alpha}$ (eq.~\eqref{eq:ssc_covariance}) are recomputed
	      at every MCMC step. This mirrors the practical strategy of freezing only the
	      expensive $S$ matrix.
	\item[\textsc{full-fix}] the entire matrix $C$ (shot noise, counts, bias and $S$)
	      is fixed.
	\item[\textsc{no-SSC}] the SSC term is dropped, retaining only the Poisson shot
	      noise (eq.~\eqref{eq:shot_noise_covariance}).
\end{description}
For \textsc{S-fix} and \textsc{full-fix}, the frozen ingredients are evaluated either at the true fiducial cosmology or at a displaced one. We define \textsc{match} as the analysis performed with the covariance built at the true cosmology; this sets the reference width $\sigma_\theta^{\rm match}$ of eq.~\eqref{eq:width_ratio}, at which \textsc{S-fix} and \textsc{full-fix} coincide.

We use \textsc{match}, rather than a fully cosmology-dependent covariance recomputed in its entirety at every MCMC step, as the reference. Our goal is to isolate the effect of evaluating the covariance at a cosmology different from that underlying the data. In our controlled study, the cosmology used to generate each mock is known exactly, so the covariance evaluated at that cosmology is, by construction, the correctly specified covariance for the corresponding data vector. A fully varying covariance differs from \textsc{match} in that, when $C$ depends on the sampled parameters, the normalization term $\ln\det C(\theta)$ in eq.~\eqref{eq:gaussian_likelihood} is no longer constant and therefore contributes to the shape of the posterior. We account for this parameter dependence in \textsc{S-fix}, as described below, so that the remaining difference between \textsc{S-fix} and a fully varying treatment is the cosmology dependence of $S_{\alpha\beta}$.

We note that the \textsc{S-fix} construction is already partially cosmology dependent, since $b_{i\alpha}\bar\mu_{i\alpha}$, and hence $C$ and $\det C$, are recomputed at every step. Thus, the parameter dependence of the normalization term is already included in \textsc{S-fix}.
We have quantified the remaining difference directly. For each of the four survey configurations, we repeated the \textsc{match} analysis with $S_{\alpha\beta}$ recomputed at every likelihood evaluation, keeping everything else unchanged: the same mock realization, the same spherical-cap footprint and redshift binning, and the same sampler settings. The marginalized uncertainties on $\Omega_{m0}$, $\sigma_8$ and $w$ then differ from their \textsc{match} counterparts by between $0.2\%$ and $1.4\%$ at $18000\,\mathrm{deg}^2$, and between $3.2\%$ and $4.9\%$ at $3000\,\mathrm{deg}^2$. The residual is larger for the smaller footprint, where the posterior is wider and the chain therefore explores a larger range of $S_{\alpha\beta}$ around its fiducial value. In both cases, this difference is well below the effects of covariance misspecification that are the subject of this work.

The fully varying treatment is, however, more expensive. Recomputing $S_{\alpha\beta}$
adds a fixed cost to each likelihood evaluation: timed single-threaded on one
machine, $0.44\,\mathrm{s}$ at $18000\,\mathrm{deg}^2$ and $1.11\,\mathrm{s}$ at
$3000\,\mathrm{deg}^2$. This overhead is the same to within $6\%$ for all three catalog
constructions of section~\ref{sec:realism}, as expected, since $S_{\alpha\beta}$
depends on neither the mass proxy nor the photometric redshifts.
The remaining cost of the likelihood evaluation, by contrast, grows sharply with realism, from
$0.32\,\mathrm{s}$ per evaluation for idealised catalogs to $1.49\,\mathrm{s}$ with
the mass proxy (mp) and $3.93\,\mathrm{s}$ with both the mass proxy and photometric redshifts
(mpz). The same overhead therefore represents a different penalty in each
case: a factor of $2.4$--$4.5$ for the idealised runs, but only $1.1$--$1.8$ for
mp and mpz. Since the idealised sweeps make up $308$ of the
$474$ analyses reported here, the penalty is largest for the bulk of this study.
We therefore recompute $S_{\alpha\beta}$ throughout only for the four reference runs
above, where the comparison is most informative; elsewhere it is held fixed, an
approximation that the agreement quoted above shows to be adequate.

\begin{figure}[ht]
	\centering
	\includegraphics[width=\textwidth]{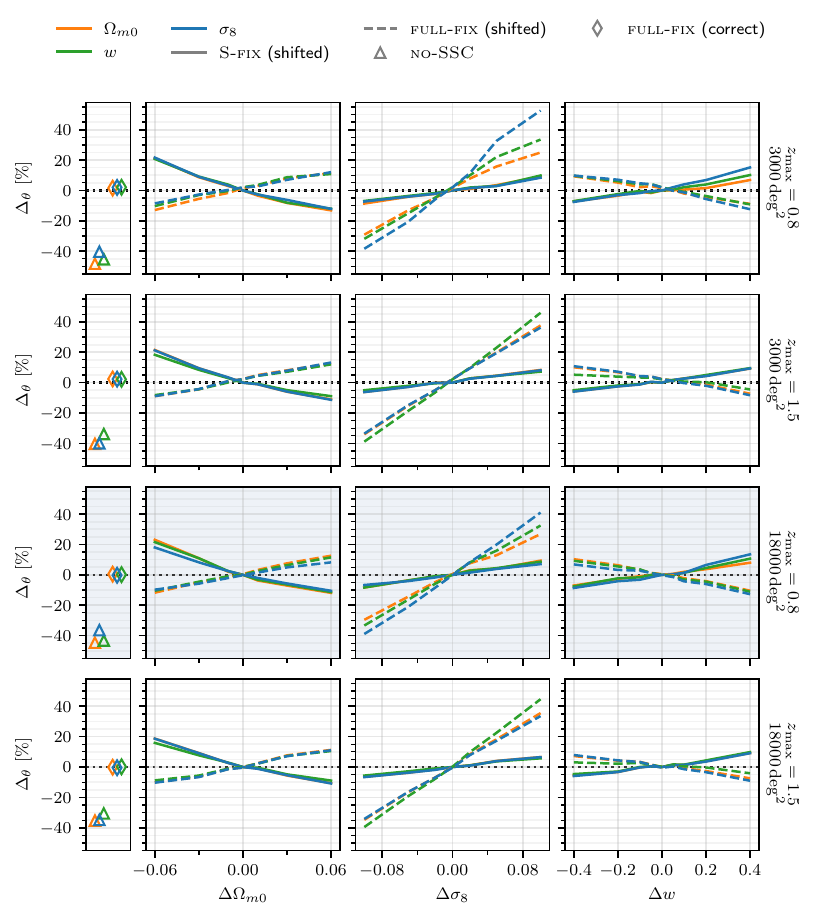}
	\caption{Relative change in the marginal posterior width,
		$\Delta_\theta=100(\sigma_\theta/\sigma_\theta^{\rm match}-1)$, for
		$\theta\in\{\Omega_{m0},\sigma_8,w\}$ as the covariance is displaced from the true
		fiducial cosmology. Rows are survey layouts; the shaded row is the layout
		quoted in the text;
		the leftmost (narrow) column shows the layout-dependent baselines (\textsc{no-SSC},
		triangles; \textsc{full-fix} at the correct cosmology, diamonds), and the
		remaining columns show the single-parameter scans in $\Omega_{m0}$, $\sigma_8$ and
		$w$. Curves are sampled at the displacements of eq.~\eqref{eq:shift_grid}.
		Solid curves (\textsc{S-fix}) fix only the $S$ matrix at the displaced cosmology, while dashed curves (\textsc{full-fix})
		freeze the full covariance.}
	\label{fig:sweep}
\end{figure}

Rather than probing only a few extreme values in parameter space, we scan each
parameter independently over a sequence of displacements,
\begin{equation}
	\label{eq:shift_grid}
	\Delta\Omega_{m0}\in\{\pm0.01,\pm0.03,\pm0.06\},\quad
	\Delta w\in\{\pm0.05,\pm0.10,\pm0.20,\pm0.40\},\quad
	\Delta\sigma_8\in\{\pm0.02,\pm0.05,\pm0.10\},
\end{equation}
and repeat the scan for both \textsc{S-fix} and \textsc{full-fix}. Each curve in figure~\ref{fig:sweep} is therefore traced by $7$--$9$ independent MCMC analyses of the same mock realization: we run a full MCMC for every shift in $\theta$, covariance construction (where applicable), and survey layout, rather than interpolating between a few points. Together with the baselines, this scan comprises $176$ complete MCMC analyses across the four survey layouts, using our default of seven redshift bins (section~\ref{sec:catalogs}). Repeating the scan with a finer binning of fourteen redshift bins, as a robustness check, for three of the four layouts brings the single-parameter total to $308$ runs.

Unless stated otherwise, we quote results for $\Omega_{\rm sky}=18000\,\mathrm{deg}^2$ and $z_{\max}=0.8$; the full sweep over all four layouts is shown in figure~\ref{fig:sweep}. As anticipated in section~\ref{sec:covariance}, the relative weight of the SSC and shot-noise terms is largely independent of the survey area, and the layout dependence of the results below is correspondingly weak: the \textsc{no-SSC} underestimation of the $\sigma_8$ width, for example, ranges from $\sim 35\%$ to $\sim 40\%$ across the four configurations. No single layout maximises the effect for every parameter -- the largest \textsc{full-fix} distortion occurs at $3000\,\mathrm{deg}^2$ for $\sigma_8$, but at $z_{\max}=1.5$ for $\Omega_{m0}$ and $w$ -- so the reference configuration used here should be read as representative rather than extremal.

Each of these $308$ configurations is a \emph{complete} MCMC analysis, not a Fisher
estimate: we use the APES sampler~\cite{d2023apes} with $500$ walkers and $120$ steps
($6\times10^4$ samples per run). The chains converge within $\sim30$--$40$ steps, and
we conservatively discard the first $50$ steps ($2.5\times10^4$ samples) as burn-in
before computing posterior statistics. Across the \emph{entire} scan, over every
displacement on $\theta$, layout and covariance construction, the location of the
posterior is remarkably stable. With the default seven redshift bins, the recovered
means depart from the reference analysis (\textsc{match}) by at most $2.2\%$ for $w$,
$0.5\%$ for $\Omega_{m0}$ and $0.25\%$ for $\sigma_8$, never more than $0.3\sigma$ of
the parameter uncertainty. The fourteen-bin repeats are slightly less stable, with
departures of up to $3.4\%$, $1.3\%$ and $0.3\%$, or $0.6\sigma$, the largest of them
in the \textsc{no-SSC} runs at $3000\,\mathrm{deg}^2$ and $z_{\max}=1.5$; in every
case the displacement remains small compared with the parameter uncertainty.
Covariance misspecification thus leaves the inferred parameters essentially unchanged
and acts almost entirely on the \emph{width} of the posterior. We therefore
characterise the effect through the width ratio $R_\theta$ in the remainder of this
section, and confirm over an ensemble of realizations that covariance misspecification adds
no detectable bias in section~\ref{sec:mc}.

Under \textsc{full-fix} at the true cosmology the posterior widths track
\textsc{match} closely, $R_\theta$ staying within $2.4\%$ of unity for every parameter
and survey layout (the diamonds in the leftmost column of figure~\ref{fig:sweep}),
an order of magnitude below the distortions discussed next. Fixing the covariance is
therefore not, by itself, the source of error: the error arises only when it is fixed
at the \emph{wrong} (displaced) cosmology.

Under \textsc{S-fix} at a displaced cosmology, with the bias and counts still
recomputed during sampling, the widths change moderately. Displacements in $\sigma_8$
have a small effect, $|\Delta_\theta|\lesssim10\%$ ($R_{\sigma_8}=0.93$ at
$\Delta\sigma_8=-0.10$, $1.07$ at $+0.10$), and those in $w$ have a comparable effect
($|\Delta_\theta|\lesssim14\%$ at $\Delta w=\pm0.40$). Displacements in $\Omega_{m0}$
are the strongest, reaching $\Delta_{\Omega_{m0}}=+23.1\%$,
$\Delta_{\sigma_8}=+18.0\%$ and $\Delta_{w}=+21.5\%$ at $\Delta\Omega_{m0}=-0.06$.
Since the $S$ matrix is the computationally expensive ingredient, this remains a useful
practical result: fixing it at an approximate cosmology, while continuing to recompute
the counts and bias at each sampled cosmology, has a substantially smaller impact than
freezing the covariance entirely, though the resulting effect is not negligible when
the assumed matter density is far from the truth.

Under \textsc{full-fix} at a displaced cosmology, by contrast, the widths distort
substantially. For $\sigma_8$ we find $R_{\sigma_8}=0.61$ at
$\Delta\sigma_8=-0.10$ and $R_{\sigma_8}=1.41$ at $+0.10$, a mis-estimation of
$\sim40\%$, with a monotonic dependence on the displacement. The effect is
controlled by the amplitude of matter fluctuations: a covariance built at a
\emph{lower} amplitude underestimates the true variance and yields artificially tight
posteriors ($R<1$), while a higher amplitude broadens them ($R>1$). Displacements in
$\Omega_{m0}$ and $w$ act in the same, weaker, sense
($R_{\sigma_8}\in[0.90,1.08]$ for $\Delta\Omega_{m0}=\pm0.06$;
$[0.87,1.07]$ for $\Delta w=\pm0.40$).

The contrast between \textsc{S-fix} and \textsc{full-fix} is informative. Their
difference is precisely whether the bias and count prefactors
$b_{i\alpha}\bar\mu_{i\alpha}$ are recomputed or fixed. The fact that \textsc{S-fix}
distorts the widths considerably less than \textsc{full-fix} shows that the
dominant cosmology dependence of the cluster-count covariance resides in the
abundance and bias prefactors, not in the $S$ matrix itself. This is physically
expected: the cluster abundance is exponentially sensitive to the fluctuation
amplitude, so holding $b\,\bar\mu$ fixed at a wrong amplitude strongly mis-scales the SSC
term, whereas $S$ varies comparatively slowly with cosmology.

The \textsc{no-SSC} construction underestimates the uncertainties substantially, and with only a weak dependence on the survey configuration. We find
$R_{\sigma_8}=0.64$ at $\Omega_{\rm sky}=18000\,\mathrm{deg}^2$ and $0.60$ at
$3000\,\mathrm{deg}^2$ for $z_{\max}=0.8$, and $0.65$ and $0.60$ respectively for
$z_{\max}=1.5$: neglecting the SSC term understates the $\sigma_8$ uncertainty by
$35$--$40\%$ in every layout considered. This near-independence of the footprint is the
expected consequence of $S_{\alpha\beta}\propto1/f_{\rm sky}$, as discussed in
section~\ref{sec:covariance}, which gives the SSC and shot-noise terms the same leading area scaling.

The fourteen-bin repeats do not change these conclusions. For the correctly specified
analysis, halving the bin width changes the marginal widths by less than $4\%$ at
$z_{\max}=1.5$ for both areas. The misspecification effects change by a few percentage
points at most; the \textsc{no-SSC} deficit, for instance, becomes $36$--$43\%$. The
only large change is at $z_{\max}=0.8$ and $3000\,\mathrm{deg}^2$, the layout with the
weakest constraints, where finer binning increases the effect: the \textsc{full-fix}
response of the $\sigma_8$ width to $\Delta\sigma_8=+0.10$ grows from $+53\%$ to
$+81\%$, and the \textsc{S-fix} response to $\Delta\Omega_{m0}=-0.06$ from $+22\%$ to
$+38\%$. Finer redshift binning therefore does not reduce the sensitivity to a
misspecified covariance, and we keep seven bins as the default.

\begin{figure}[!ht]
	\centering
	\includegraphics[width=0.49\textwidth]{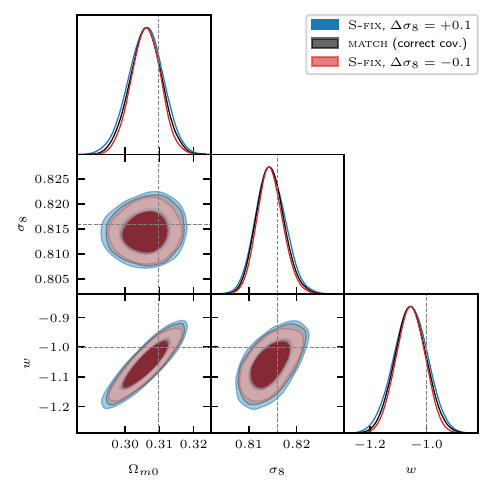}
	\includegraphics[width=0.49\textwidth]{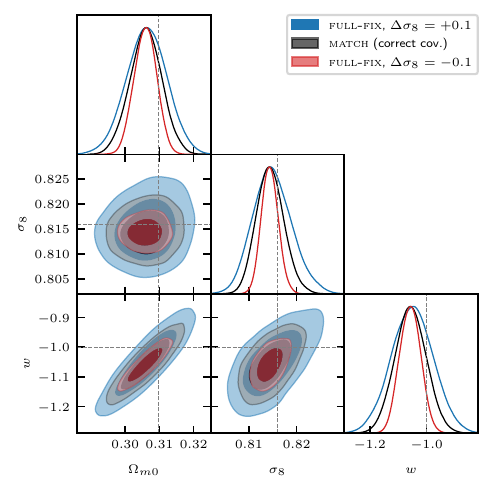}
	\caption{Posterior constraints on $(\Omega_{m0},\sigma_8,w)$ from a single mock
		realization for the representative survey configuration ($\Omega_{\rm
				sky}=18000\,\mathrm{deg}^2$, $z_{\max}=0.8$). The left panel compares the
		reference analysis (\textsc{match}, black) with analyses in which only the $S$
		matrix is fixed at cosmologies with $\Delta\sigma_8=\pm0.10$ (blue and red).
		The right panel shows the corresponding comparison when the entire covariance
		is fixed at those displaced cosmologies. Contours denote the $68\%$ and $95\%$
		credible regions, and dashed lines indicate the fiducial parameter values. Both
		panels are plotted using identical parameter ranges, allowing a direct
		comparison of posterior widths. Fixing only $S$ has a negligible impact on the
		inferred constraints, with contours nearly coincident with the reference
		result. In contrast, fixing the full covariance produces a substantially larger
		change in the posterior volume, broadening or narrowing the constraints
		depending on the sign of the covariance mismatch, while leaving the posterior
		centre essentially unchanged.}
	\label{fig:corner:sigma8}
\end{figure}

\begin{figure}[!ht]
	\centering
	\includegraphics[width=\textwidth]{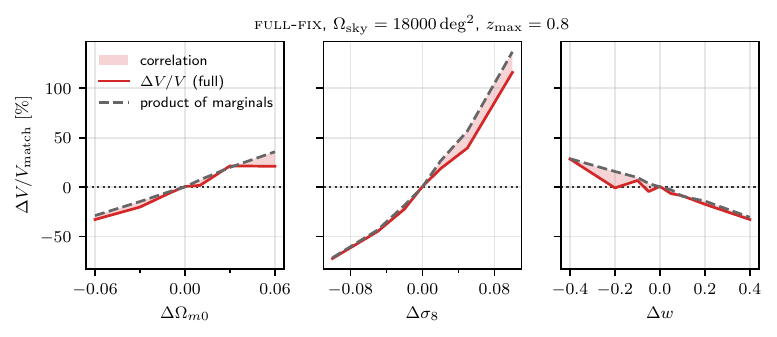}
	\caption{Joint $68\%$ credible-volume change $\Delta V/V_{\rm match}$ at the representative
		layout ($\Omega_{\rm sky}=18000\,\mathrm{deg}^2$, $z_{\max}=0.8$) with the full
		covariance fixed at a displaced cosmology, as a function of the displacement in
		$\Omega_{m0}$, $\sigma_8$ and $w$. The solid curve is the joint volume; the dashed
		curve is the product of the three marginal width ratios (the value expected if the
		correlations were unchanged). The shaded gap is the contribution of the altered
		correlation/degeneracy structure, which the per-parameter widths do not capture.}
	\label{fig:volume}
\end{figure}

Figure~\ref{fig:corner:sigma8} shows these effects directly for two representative
cases, overlaying the posterior of $(\Omega_{m0},\sigma_8,w)$ from the same mock under
\textsc{match}, \textsc{S-fix} at $\Delta\sigma_8=\pm0.10$, and \textsc{full-fix} at
the same displaced cosmologies. The \textsc{S-fix} posteriors (left panel) are
essentially identical to \textsc{match}, whereas the \textsc{full-fix} ones (right
panel) are visibly broadened or narrowed depending on the sign of the covariance
mismatch; all five share the same location, marked by the fiducial values (consistent
with the absence of bias established in section~\ref{sec:mc} but with a natural
statistical shift). The figure therefore illustrates the central result of this
section: covariance misspecification affects primarily the size of the credible
regions, together with their shape and the parameter correlations, both of which are
quantified through the joint credible volume of figure~\ref{fig:volume}, rather than
the location of the posterior maximum.

\begin{figure}[!ht]
	\centering
	\includegraphics[width=0.49\textwidth]{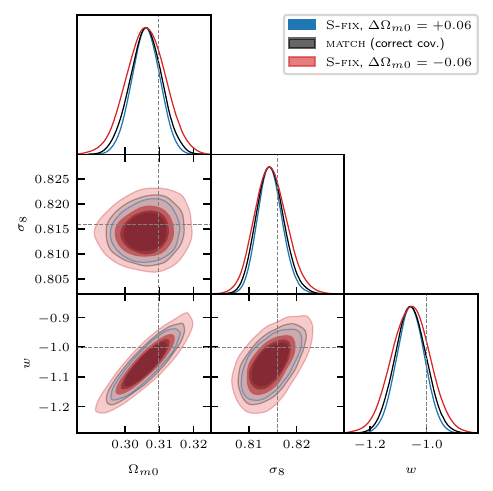}
	\includegraphics[width=0.49\textwidth]{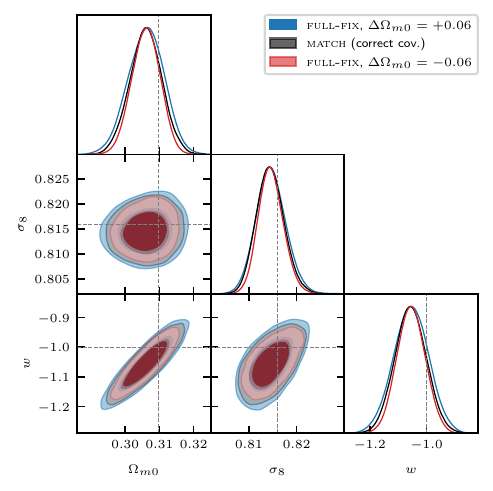}
	\caption{Same as figure~\ref{fig:corner:sigma8}, but for covariance mismatches
		generated by displacing the fiducial matter density by
		$\Delta\Omega_{m0}=\pm0.06$. The left panel compares the \textsc{S-fix}
		prescription, in which only the $S$ matrix is fixed at the displaced cosmology,
		while the right panel shows the \textsc{full-fix} prescription, in which the
		entire covariance is fixed. Black contours correspond to the reference
		(\textsc{match}) analysis, and blue and red contours to positive and negative
		parameter displacements, respectively. Contours denote the $68\%$ and $95\%$
		credible regions, dashed lines indicate the fiducial parameter values, and both
		panels use identical parameter ranges. In contrast to
		figure~\ref{fig:corner:sigma8}, the impact of covariance misspecification is of
		comparable magnitude for the \textsc{S-fix} and \textsc{full-fix}
		prescriptions, but with opposite trends: positive and negative displacements
		broaden different sides of the comparison. This reflects, in the full posterior
		distribution, the same behaviour observed in figure~\ref{fig:sweep}.}
	\label{fig:corner:omega_m}
\end{figure}

Figure~\ref{fig:corner:omega_m} presents the analogous comparison for covariance
mismatches generated by displacing $\Omega_{m0}$. In contrast to the $\sigma_8$ case,
the impact of covariance misspecification is of comparable magnitude for the
\textsc{S-fix} and \textsc{full-fix} prescriptions, but with opposite trends: a
displacement that broadens the posterior in one prescription narrows it in the other.
Together, figures~\ref{fig:corner:sigma8} and~\ref{fig:corner:omega_m} provide the full
posterior view of the behaviour quantified in figure~\ref{fig:sweep}, showing how the
changes in variance ratios translate into the geometry of the multidimensional
credible regions.

While the per-parameter ratios $R_\theta$ summarise the marginal widths, it is useful to
also quote a single joint figure of merit. For this we use the volume of the
highest-posterior-density (HPD) credible region in the three-dimensional
$(\Omega_{m0},\sigma_8,w)$ space, that is, the smallest region enclosing a fixed fraction
$\alpha$ of the posterior; we take $\alpha=0.6827$ (the $68\%$ region), so that a smaller
volume corresponds to a tighter joint constraint. We estimate it directly from the MCMC
chain with the procedure implemented in \texttt{NumCosmo}, which selects the $\alpha$
fraction of samples with the highest posterior probability and evaluates the enclosed
parameter-space volume (appendix~\ref{app:HDP}). Denoting this volume $V$, we report its
change relative to the \textsc{match} analysis, $\Delta V/V_{\rm match}$. Being a joint
measure, $V$ is complementary to the marginals in
two respects. First, it compounds the individual width changes: for a Gaussian posterior
$V=\big(\prod_i\sigma_i\big)\sqrt{\det R}$, with $R$ the correlation matrix, so
$\Delta V/V$ is of the order of the sum of the per-parameter $\Delta_\theta$ and is
therefore much larger than any single one. For \textsc{full-fix} at
$\Delta\sigma_8=+0.10$ (representative layout) the credible volume more than doubles
($\Delta V/V_{\rm match}\simeq+116\%$), against the $\sim41\%$ change of the $\sigma_8$
marginal alone; at $\Delta\sigma_8=-0.10$ it shrinks by $\sim72\%$, while \textsc{S-fix}
remains mild ($\simeq+13\%$). Second, and unlike the $R_\theta$, $V$ responds
to changes in the correlation structure $R$ itself. Figure~\ref{fig:volume} isolates this
contribution: the solid curve is $\Delta V/V$ and the dashed curve is the product of the
marginal width ratios, that is, the value $V$ would take if the correlations were
unchanged; the shaded difference between them is the effect of the shifting degeneracy
directions. A fixed covariance therefore misestimates not only the size of the credible
region but also its shape, an effect that only a joint measure such as $V$ makes explicit.

\subsection{Combined parameter shifts}
\label{sec:combined}

\begin{figure}[!ht]
	\centering
	\includegraphics[width=0.6\textwidth]{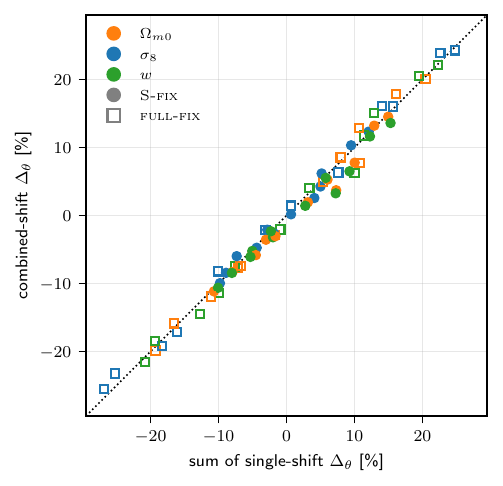}
	\caption{Test for non-linear parameter coupling at the representative layout
		($\Omega_{\rm sky}=18000\,\mathrm{deg}^2$, $z_{\max}=0.8$). For each two-parameter
		shift, the combined width change $\Delta_\theta$ is plotted against the sum of the
		two single-parameter changes, for $\Omega_{m0}$, $\sigma_8$ and $w$ and for
		\textsc{S-fix} (filled) and \textsc{full-fix} (open). Points
		follow the identity line (dotted), confirming that the shifts superpose linearly.}
	\label{fig:combined}
\end{figure}

The analysis above displaces one parameter at a time. To check for non-linear
responses, we repeat it for simultaneous two-parameter displacements (in
$\Omega_{m0}\times w$, $\Omega_{m0}\times\sigma_8$ and $w\times\sigma_8$) at the
representative layout, under both \textsc{S-fix} and \textsc{full-fix}, for a total of $24$
MCMC analyses. In every case the
combined effect equals the sum of the individual effects to within the realization
noise: for example, under \textsc{full-fix},
$(\Delta\Omega_{m0},\Delta\sigma_8)=(+0.03,+0.05)$ yields
$\Delta_{\sigma_8}=+24.3\%$ against a sum of individual shifts of $+24.8\%$, and
$(\Delta w,\Delta\sigma_8)=(-0.10,-0.05)$ yields $-19.2\%$ against $-18.3\%$. Across
all $24$ combined runs the joint effect never departs from the sum of the two
individual effects by more than $4.0$ percentage points, with a median departure of
$0.8$, to be compared with combined effects reaching $25\%$. No non-linear coupling
between parameters is detected, so the single-parameter sweeps of
section~\ref{sec:s_matrix} fully characterise the behaviour over the relevant region of
parameter space. Figure~\ref{fig:combined} makes this explicit: the combined-shift
width change tracks the sum of the corresponding single-shift changes along the
identity line for all parameters and for both \textsc{S-fix} and \textsc{full-fix}.

\subsection{Consistency of the forward and mirror configurations}
\label{sec:mirror}

In this section, we compare the two complementary analyses introduced in section~\ref{sec:results}. In the \emph{forward} configuration, the mock catalogue is kept fixed while the covariance is constructed at displaced cosmologies. In the \emph{mirror} configuration, the covariance is instead fixed while the mock catalogue is generated at displaced cosmologies.

The \emph{forward} and \emph{mirror} configurations differ only in which cosmology is assigned to the mock data and which to the covariance matrix. Since our interpretation is that the effect of covariance misspecification is controlled by the mismatch between these two cosmologies, both configurations should produce the same results when compared at equal cosmological offsets. Although this expectation is physically well motivated, it is not guaranteed a priori in a nonlinear MCMC analysis with a cosmology-dependent covariance. In this section, we verify this expectation explicitly.

For this, we reanalyse the representative layout in the \emph{mirror}
configuration, generating the mock at a displaced cosmology
($\Delta\Omega_{m0}=\pm0.03$, $\Delta\sigma_8=\pm0.05$, $\Delta w=\pm0.10$) and fitting it
with the covariance held at the fiducial value, under both \textsc{S-fix} and
\textsc{full-fix}. With the matching correct-covariance analyses at each displaced cosmology,
this amounts to $24$ MCMC analyses. Each \emph{mirror} run is normalised to its own \textsc{match}
reference, the analysis with the covariance built consistently at that displaced
cosmology, so that $\Delta_\theta$ again measures the misspecified width against the
correct one.

\begin{figure}[!ht]
	\centering
	\includegraphics[width=\textwidth]{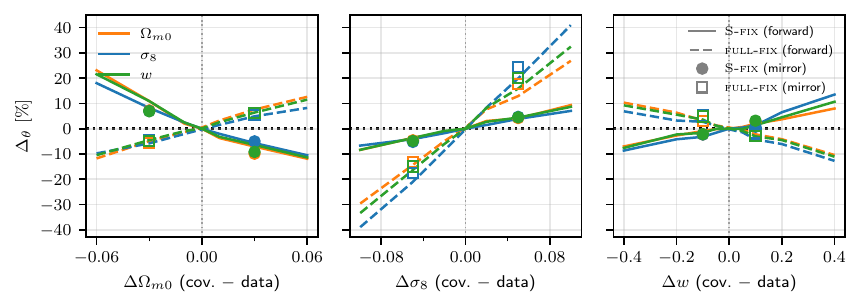}
	\caption{Equivalence of the forward and mirror configurations at the representative layout
		($\Omega_{\rm sky}=18000\,\mathrm{deg}^2$, $z_{\max}=0.8$), shown on a common
		cosmology-mismatch axis $\delta\theta=\theta_{\rm cov}-\theta_{\rm data}$. Curves are the
		forward single-parameter sweeps (solid: \textsc{S-fix}; dashed: \textsc{full-fix});
		markers are the mirror runs (filled circles: \textsc{S-fix}; open squares:
		\textsc{full-fix}), generated at a displaced cosmology and plotted at
		$\delta\theta=-\mathrm{shift}$. Colours denote the inferred parameter. The mirror points
		lie on the forward curves, confirming that the effect is set by the data-covariance
		mismatch alone.}
	\label{fig:mirror}
\end{figure}

The key observation is that what drives the effect is the signed mismatch between the
covariance and the data cosmologies, $\delta\theta\equiv\theta_{\rm cov}-\theta_{\rm data}$, where
$\theta_{\rm data}$ is the cosmology used to generate the mock and $\theta_{\rm cov}$ the
cosmology at which the frozen part of the covariance (the full matrix under \textsc{full-fix},
the $S$ matrix alone under \textsc{S-fix}) is evaluated. In
the \emph{forward} sweep a covariance displaced by $+s$ on fiducial data gives $\delta\theta=+s$;
in the \emph{mirror} runs a covariance held at the fiducial value while the data move by
$+t$ gives $\delta\theta=-t$. Placing both on this common axis, figure~\ref{fig:mirror} shows the
\emph{mirror} points falling onto the \emph{forward} curves for every inferred parameter
and for both covariance constructions. For instance, freezing the full covariance under a
$\sigma_8$ mismatch of $+0.05$ inflates $\sigma_{\sigma_8}$ by $+20\%$ in the \emph{forward}
sweep and by $+24\%$ in the \emph{mirror} run, while \textsc{S-fix} leaves it within a few
percent in both cases. Across all $36$ comparisons at matched cosmology mismatch (six
displacements, two covariance constructions and three inferred parameters) the forward
and mirror results differ by a median of $1.1$ and at most $4.7$ percentage points,
consistent with the realization noise of single mocks. The impact of covariance misspecification therefore depends only on the cosmology mismatch, and not on which of the data or the covariance is displaced, so the \emph{forward} scan of section~\ref{sec:s_matrix} is representative of a real analysis.

\subsection{Bias of the parameter estimators}
\label{sec:mc}

The results above concern the \emph{width} of the posterior on a single realization; we now
turn to the \emph{location} of the estimators. Point estimators such as the posterior mean,
or equivalently the maximum-likelihood estimate, are in general only \emph{asymptotically} unbiased: for a finite data sample they can carry an intrinsic bias that is present even when the covariance is built correctly. Whether this intrinsic bias is present in our analysis cannot be inferred from the Gaussianity of the likelihood alone. Although the mock data are drawn from eq.~\eqref{eq:gaussian_likelihood}, the Gaussianity applies to the data vector, whereas the best-fit cosmological parameters are obtained through the nonlinear model prediction $\mu(\theta)$. The estimator is therefore a nonlinear function of the data. Consequently,  covariance misspecification could in principle add to  this intrinsic bias.
The relevant question is therefore not whether the estimators are exactly unbiased, but
whether misspecifying the covariance \emph{introduces} a bias beyond the intrinsic one. To
answer it we measure the \emph{total} estimator bias in each configuration and include, as a
control, the correct-covariance case with no cosmology shift, against which the misspecified
cases can be compared.

We therefore perform a Monte Carlo study at the representative layout, drawing
$N=1000$ independent mock catalogs per configuration.  For each realization, we determine the maximum-likelihood (best-fit)
cosmological parameters by minimizing $-2\ln\mathcal{L}_{\rm G}$, with
$\mathcal{L}_{\rm G}$ the Gaussian likelihood of eq.~\eqref{eq:gaussian_likelihood},
with respect to the three free cosmological parameters $(\Omega_{m0}, \sigma_8, w)$.
The minimization is performed with the NLOpt\footnote{\url{https://nlopt.readthedocs.io}}
optimization framework interfaced via \texttt{NumCosmo}.

The configurations are arranged so that every misspecified analysis can be compared against a correctly specified one \emph{at the same true cosmology}. In the \emph{forward} configuration the mocks are generated at the fiducial cosmology and analysed either with the correct covariance
(\textsc{match}) or with the SSC term dropped (\textsc{no-SSC}). In the \emph{mirror}
configuration the mocks are generated at each displaced cosmology and analysed either with
the covariance rebuilt consistently at that same cosmology (the correct, \textsc{match},
analysis) or with the covariance held at the fiducial value (the misspecified analysis). Each of these is run for both freezings, \textsc{S-fix} and
\textsc{full-fix}. This yields two groups: a \emph{correct-covariance} control (the forward
\textsc{match} run and the mirror runs analysed at their own cosmology) and a
\emph{misspecified} set (the forward \textsc{no-SSC} run and the mirror runs analysed with
the covariance fixed at the fiducial value). Comparing the two groups at matched true
cosmology isolates any bias introduced by misspecification from the intrinsic
finite-sample bias common to both. In total this covers $28$ covariance configurations,
each resampled $N=1000$ times, for  a
total of $28{,}000$ independent minimizations of $-2\ln\mathcal{L}_{\rm G}$, each
yielding a best-fit vector for the three cosmological parameters
$(\Omega_{m0}, \sigma_8, w)$.

For each mock we record $\hat\theta$, the best-fit (point) estimate of the parameter
$\theta$ returned by the fit. We define the estimator bias as
$\mathrm{Bias}(\hat\theta)=\langle\hat\theta\rangle-\theta_{\rm true}$, where
$\langle\hat\theta\rangle$ is the mean of $\hat\theta$ over the ensemble and
$\theta_{\rm true}$ is the value of that parameter used to generate the mocks, that is, the
resample cosmology at which the catalogs are drawn. We quote the bias in units of the
per-realization scatter $\sigma_\theta$, the standard deviation of $\hat\theta$ across the
ensemble. Since $\langle\hat\theta\rangle$ is the mean of $N$ independent estimates,
its standard error is $\sigma_\theta/\sqrt{N}$, so the ensemble resolves a bias only down to
$1/\sqrt{N}\approx0.03\,\sigma$; under the null hypothesis of no bias,
$\mathrm{Bias}(\hat\theta)/\sigma_\theta$ has standard deviation $1/\sqrt{N}$ and should fall
within $\pm2/\sqrt{N}$, two standard errors of the mean, about $95\%$ of the time.

\begin{figure}[!ht]
	\centering
	\includegraphics[width=\textwidth]{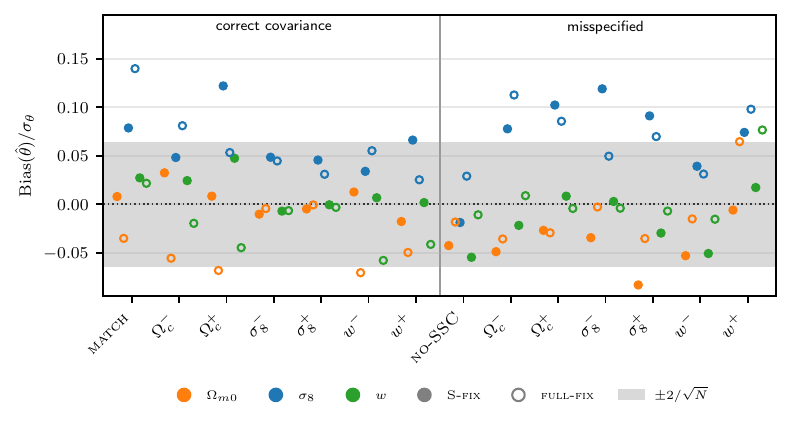}
	\caption{Estimator bias in units of the parameter scatter,
		$\mathrm{Bias}(\hat\theta)/\sigma_\theta$, from the $N=1000$ Monte Carlo ensemble at
		the representative layout ($\Omega_{\rm sky}=18000\,\mathrm{deg}^2$, $z_{\max}=0.8$).
		Colour denotes the inferred parameter ($\Omega_{m0}$, $\sigma_8$, $w$) and the marker
		fill the covariance freezing (filled: \textsc{S-fix}; open: \textsc{full-fix}); each
		column is one configuration. Columns are split into the \emph{correct-covariance}
		control, where the covariance is built at the cosmology that generated the mock, and the
		\emph{misspecified} set. The control comprises the \emph{forward} \textsc{match} run
		(mock and covariance at the fiducial cosmology) and the \emph{mirror} runs at each
		shift (mock at the displaced cosmology, covariance rebuilt there). The misspecified set
		comprises the \emph{forward} \textsc{no-SSC} run (SSC term dropped on a fiducial mock)
		and the \emph{mirror} runs at each shift with the covariance held at the fiducial value
		($\Omega_c$, $\sigma_8$, $w$ shifted by $\pm$). The shaded band is
		$\pm2/\sqrt{N}$, two standard errors of the ensemble mean (about $95\%$ C.L.) expected for
		an unbiased estimator. The two groups scatter identically within the band, showing that
		covariance misspecification does not displace the recovered parameters.}
	\label{fig:mc_bias}
\end{figure}

No bias is added by misspecifying the covariance (figure~\ref{fig:mc_bias}); the only bias
resolved is a small intrinsic offset present in all configurations. Already the
correct-covariance controls carry $|\mathrm{Bias}(\hat\theta)|\le0.14\,\sigma$ (median
$0.03\,\sigma$; in absolute terms
$3\times10^{-6}\lesssim|\mathrm{Bias}(\hat\theta)|\lesssim3\times10^{-3}$). The
misspecified set scatters within the same band and is statistically indistinguishable from
the controls at the matched true cosmology, whether the covariance is stripped of its SSC
term on a fiducial mock (\emph{forward} \textsc{no-SSC}) or held at the fiducial value while
the data are displaced (the \emph{mirror} case of section~\ref{sec:mirror}).
Of the $84$ parameter estimates, those for $\Omega_{m0}$ and $w$ fall within the
$\pm2/\sqrt{N}$ band at close to the expected rate, only $2$ of $14$ $\Omega_{m0}$ estimates and none of the $14$ $w$ estimates lie outside it in the control group, compared with $2$ and $1$ of $14$, respectively, in the misspecified group, with
no systematic trend with the shift. The $\sigma_8$ estimates behave differently:
across the ensembles $\sigma_8$ is recovered marginally high and $\Omega_{m0}$ marginally low (mean biases $+0.066\,\sigma$ and
$-0.022\,\sigma$, of consistent sign across the configurations), the opposite signs tracing
the $\Omega_{m0}$--$\sigma_8$ anti-correlation, i.e.\ a small displacement along the
degeneracy direction. Because this $\sigma_8$ offset sits essentially on the band edge,
$2/\sqrt{N}=0.063\,\sigma$, the $\sigma_8$ estimates fall outside the band in $5$ of $14$
control and $9$ of $14$ misspecified configurations; the difference between the two groups
is not statistically significant (Fisher exact test, $p=0.26$). This coherent offset is the
intrinsic finite-sample bias of the estimator: it stays below $0.15\,\sigma$ and, crucially,
is the same in the correct-covariance controls as in the misspecified configurations, so it
is not induced by the covariance choice. Covariance misspecification therefore adds no
detectable bias; its effect is confined to the inferred uncertainties, as quantified in
section~\ref{sec:s_matrix}, not the location of the recovered parameters. The Monte
Carlo ensemble also provides a second reading of the credible volume of
section~\ref{sec:s_matrix}: the covariance volume of the best-fit ensemble, computed as
$\sqrt{\det\mathrm{Cov}}$ over the $N=1000$ best fits, agrees with the HPD volume of the
corresponding MCMC posterior. The same joint volume can thus be obtained either from the
posterior chains or from the ensemble of estimators (appendix~\ref{app:HDP}),
confirming that the volume estimates used in this work faithfully represent the
statistical scatter of the estimators.

\subsection{Observational realism: mass proxy and photometric redshifts}
\label{sec:realism}

The analyses so far use idealised catalogs with true masses and redshifts. We now repeat
the representative forward sweep with two more realistic catalog constructions
(section~\ref{sec:catalogs}): mp, which replaces the true mass by the richness
through the mass-observable relation (four richness bins over $\lambda\in[20,200]$), and
mpz, which additionally replaces the true redshift by a Gaussian photometric
estimate with an LSST-like scatter $\sigma^{\rm ph}_0=0.03$. Each realism level is analysed
against its own \textsc{match} reference, so the width ratios continue to isolate the
covariance choice. Each level follows the representative single-parameter sweep of
section~\ref{sec:s_matrix} under both freezings ($44$ MCMC analyses) together with the
match-family Monte Carlo ($4{,}000$ best fits).

\begin{figure}[!ht]
	\centering
	\includegraphics{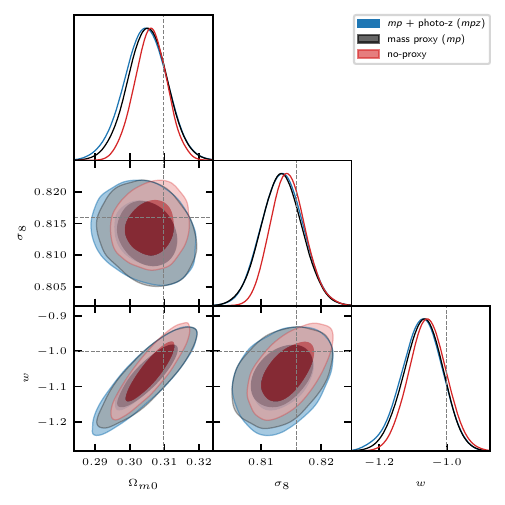}
	\caption{Correctly-specified (\textsc{match}) posterior of
		$(\Omega_{m0},\sigma_8,w)$ at the representative layout for the three realism
		levels: true masses and redshifts (no-proxy, red), mass proxy (mp,
		grey), and mass proxy plus LSST-like photo-z (mpz, blue). Contours are
		the $68\%$ and $95\%$ credible regions; dashed lines mark the fiducial values.
		The constraints broaden with realism while preserving their orientation and closely the central values; mpz is barely wider than mp, confirming that LSST-like photo-z adds little beyond the mass proxy.}
	\label{fig:realism_corner}
\end{figure}

Adding observational scatter degrades the constraints, as expected. Table~\ref{tab:realism_constraints}
lists the absolute marginal $1\sigma$ errors of the correctly-specified analysis: at the true
cosmology $\sigma_8$ is constrained to $0.36\%$, $\Omega_{m0}$ to $1.5\%$ and $w$ to $5.6\%$
for idealised catalogs, degrading to $0.42\%$, $1.9\%$ and $5.9\%$ for mp and to
$0.42\%$, $2.0\%$ and $6.2\%$ for mpz. In other words the widths broaden by
$24\%$ on $\Omega_{m0}$, $17\%$ on $\sigma_8$ and $5\%$ on $w$ for mp, and by
$32\%$, $18\%$ and $11\%$ for mpz: the realistic photo-z scatter adds little to the
marginal $\sigma_8$ constraint beyond the mass proxy, and only a modest amount on
$\Omega_{m0}$ and $w$. These penalties are modest because the SSC contribution already dominates the
error budget of the idealised catalogs: observational scatter is then added on top of a
floor that is largely set by sample variance rather than by shot noise. Had the
super-sample term been evaluated in the full-sky limit, which underestimates
$S_{\alpha\beta}$ for any finite footprint (section~\ref{sec:covariance}), that floor
would be lower and the same scatter would degrade the constraints considerably more.
Figure~\ref{fig:realism_corner} shows the same posteriors in the
$(\Omega_{m0},\sigma_8,w)$ planes: the contours broaden from \emph{true} through mp to
mpz while retaining the same orientation, with mpz only marginally wider than mp.
The recovered means lie within the single-realization scatter (within about $1\sigma$) of
the fiducial at every realism level, consistent with the unbiasedness established in
section~\ref{sec:mc}.

\begin{table}[!ht]
	\centering
	\caption{Marginal $1\sigma$ constraints (posterior standard deviations) from the
		correctly-specified (\textsc{match}) analysis at the representative layout
		($\Omega_{\rm sky}=18000\,\mathrm{deg}^2$, $z_{\max}=0.8$), for the three realism
		levels: true masses and redshifts, mass proxy (mp), and mass proxy plus
		LSST-like photo-z (mpz). Values in parentheses are the fractional constraint
		$\sigma_\theta/|\theta_{\rm fid}|$, with
		$(\Omega_{m0},\sigma_8,w)_{\rm fid}=(0.3098,0.8159,-1)$.}
	\label{tab:realism_constraints}
	\begin{tabular}{lccc}
		\hline
		Parameter              & true               & mp                 & mpz                \\
		\hline
		$\sigma_{\Omega_{m0}}$ & $0.0047\ (1.5\%)$  & $0.0058\ (1.9\%)$  & $0.0062\ (2.0\%)$  \\
		$\sigma_{\sigma_8}$    & $0.0029\ (0.36\%)$ & $0.0034\ (0.42\%)$ & $0.0034\ (0.42\%)$ \\
		$\sigma_w$             & $0.056\ (5.6\%)$   & $0.059\ (5.9\%)$   & $0.062\ (6.2\%)$   \\
		\hline
	\end{tabular}
\end{table}

\begin{figure}[!ht]
	\centering
	\includegraphics[width=\textwidth]{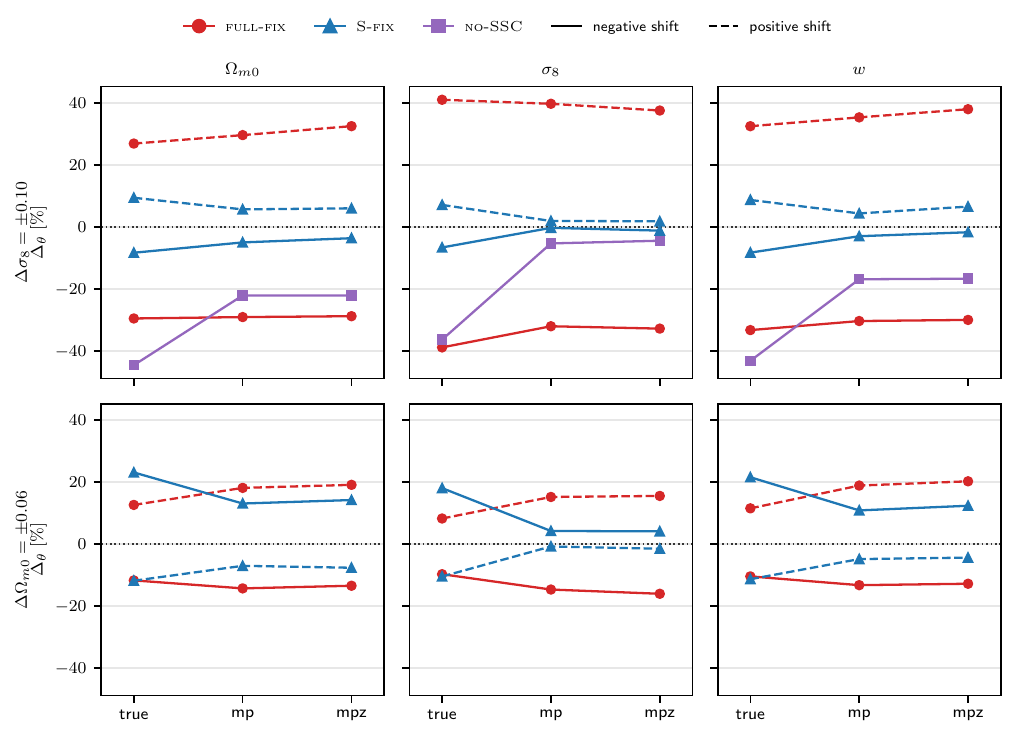}
	\caption{Robustness of the covariance-misspecification effect to observational realism
		at the representative layout ($\Omega_{\rm sky}=18000\,\mathrm{deg}^2$, $z_{\max}=0.8$).
		Columns are the inferred parameters. The upper row follows the $\sigma_8$
		displacement, $\Delta\sigma_8=\pm0.10$, together with the \textsc{no-SSC} limit; the
		lower row follows the $\Omega_{m0}$ displacement, $\Delta\Omega_{m0}=\pm0.06$, the
		strongest \textsc{S-fix} response of section~\ref{sec:s_matrix}. Colours and markers
		denote the covariance construction (\textsc{full-fix}, red circles; \textsc{S-fix},
		blue triangles; \textsc{no-SSC}, purple squares); solid (dashed) lines are negative
		(positive) displacements. The width change
		$\Delta_\theta=100(\sigma_\theta/\sigma_\theta^{\rm match}-1)$ is plotted against
		the realism level (true masses and redshifts; mass proxy, mp; mass proxy plus
		LSST-like photo-z, mpz), normalised to the \textsc{match} reference of the same
		realism. The \textsc{full-fix} effect remains large and roughly constant with
		realism. The \textsc{S-fix} effect is below $7\%$ for $\sigma_8$ displacements at the
		two proxy levels, and $13$--$14\%$ on $\Omega_{m0}$ for $\Omega_{m0}$ displacements. The \textsc{no-SSC} underestimation shrinks
		substantially as the proxy scatter lowers the relative weight of the SSC term.}
	\label{fig:realism}
\end{figure}

The impact of covariance misspecification, by contrast, is essentially unchanged across
realism (figure~\ref{fig:realism}). The dominant \textsc{full-fix} distortion persists
at both proxy levels: at $\Delta\sigma_8=+0.10$ ($-0.10$) the $\sigma_8$ width is
mis-estimated by $+41\%$ ($-39\%$) for true masses, $+40\%$ ($-32\%$) for mp and
$+38\%$ ($-33\%$) for mpz, and the joint credible volume more than doubles in
every case ($\Delta V/V_{\rm match}\simeq+116\%$, $+139\%$, $+134\%$ at
$\Delta\sigma_8=+0.10$) and shrinks to roughly a third of it ($0.28$--$0.34\,V_{\rm
			match}$) at $\Delta\sigma_8=-0.10$. The
\textsc{S-fix} construction has a small effect for the $\sigma_8$ displacements at all
realism levels ($|\Delta_\theta|\lesssim10\%$ for true masses and $\lesssim7\%$ at the
two proxy levels). Its largest effect, for the $\Omega_{m0}$ displacement
(section~\ref{sec:s_matrix}; lower row of figure~\ref{fig:realism}), decreases with
realism but remains above $10\%$: at $\Delta\Omega_{m0}=-0.06$ the $\Omega_{m0}$ width
is overestimated by $+23\%$ for true masses and by $+13\%$ (mp) and $+14\%$ (mpz); the
$w$ width behaves in the same way, while the $\sigma_8$ width overestimate drops from
$+18\%$ to $+4\%$. The \textsc{full-fix} effect for the same displacement does not
decrease; on $\Omega_{m0}$ it grows from $\pm12\%$ to $\pm14$--$19\%$. The other qualitative change is
in the \textsc{no-SSC} limit, whose underestimation shrinks markedly with realism: the $\sigma_8$
($\Omega_{m0}$) error is underestimated by $36\%$ ($45\%$) for true masses but only by
$\sim5\%$ ($\sim22\%$) at both mp and mpz. This is consistent with the
physical picture of section~\ref{sec:s_matrix}: observational scatter raises the
shot-noise floor and so lowers the relative weight of the SSC term, which makes
dropping or freezing $S$ less consequential, while the \textsc{full-fix} effect, driven
by the abundance and bias prefactors rather than by $S$, is essentially
realism-independent.

\begin{figure}[!ht]
	\centering
	\includegraphics[width=\textwidth]{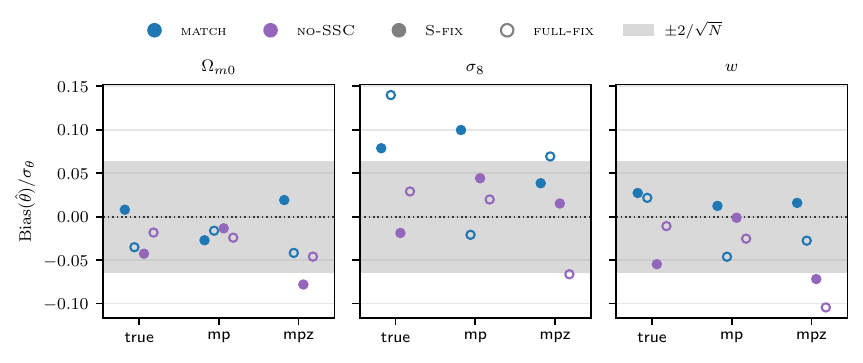}
	\caption{Estimator bias in units of the parameter scatter,
		$\mathrm{Bias}(\hat\theta)/\sigma_\theta$, from the Monte Carlo analyses across the
		three realism levels (true masses and redshifts; mass proxy, mp; mass proxy
		plus LSST-like photo-z, mpz), with $N=1000$ realizations each, for the representative layout
		($\Omega_{\rm sky}=18000\,\mathrm{deg}^2$, $z_{\max}=0.8$). Each panel corresponds to an inferred parameter. For each realism level, the forward match-family controls are shown:
		\textsc{match} (correct covariance) and \textsc{no-SSC}, each with the \textsc{S-fix}
		(filled) and \textsc{full-fix} (open) freezings. The shaded band is $\pm2/\sqrt{N}$, corresponding to two standard errors of the mean ($\sim95\%$ C.L.). The points scatter around zero within, or marginally outside, their bands at every realism level, and the small intrinsic offset is consistent with that seen for true masses, indicating that
		observational realism neither introduces nor amplifies an estimator bias. The proxy
		Monte Carlo covers the forward match family only; the shifted cases are covered for true masses in figure~\ref{fig:mc_bias}.}
	\label{fig:realism_bias}
\end{figure}

The estimator-bias study of section~\ref{sec:mc}, repeated at the mp and mpz
realism levels (the match-family controls, $N=1000$ realizations), confirms that the estimators
remain unbiased there as well (figure~\ref{fig:realism_bias}): the proxy biases stay
below $0.11\,\sigma$, within the $0.15\,\sigma$ envelope of the idealised ensembles of
section~\ref{sec:mc}, and the same small intrinsic offset noted there, with
$\Omega_{m0}$ recovered marginally low, persists unchanged
across realism. The results obtained with idealised catalogs are therefore robust to a
realistic mass proxy and to LSST-like photometric redshifts.

\begin{figure}[!ht]
	\centering

	\includegraphics[width=0.48\linewidth]{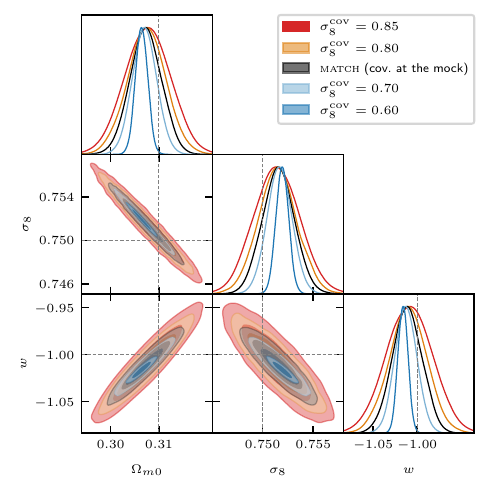}
	\includegraphics[width=0.48\linewidth]{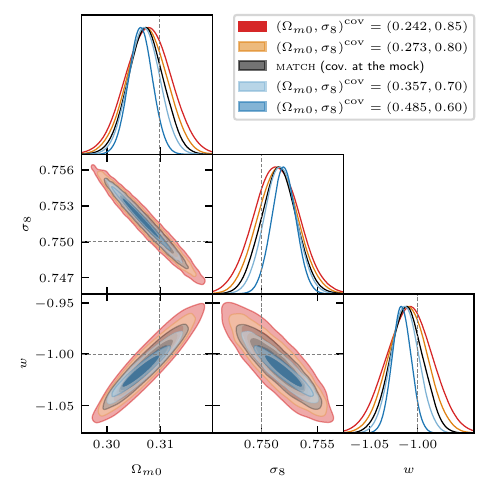}\\
	\includegraphics[width=0.48\linewidth]{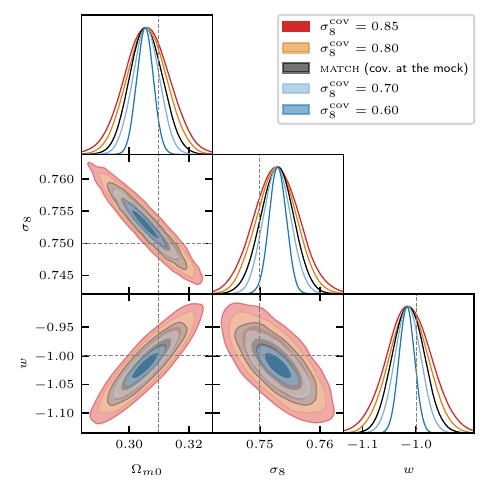}
	\includegraphics[width=0.48\linewidth]{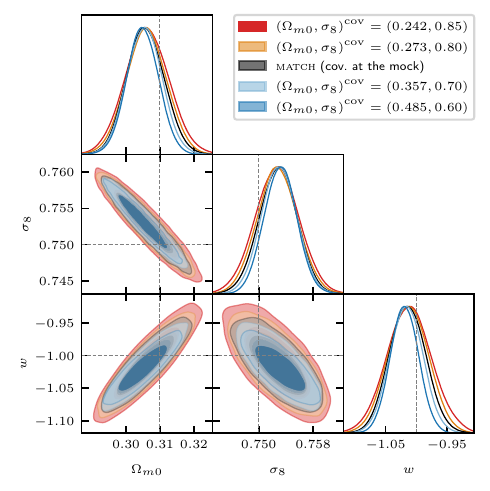}

	\caption{Corner plots for the cosmological parameters $\{\Omega_{m0}, \sigma_8,
			w\}$, with sky coverage $\Omega_{\mathrm{sky}}=18000 \,\mathrm{deg}^2$ and
		maximum redshift $z_\text{max}=1.5$. Upper (lower) panels correspond to
		analyses assuming no-proxy (mpz). Contours denote the 68\% and 95\%
		credible regions of the parameter posteriors. Left (right) panels show
		results obtained from MCMC when varying $S_8$ ($S_8$ fixed). Grey contours
		correspond to analyses in which the covariance matrix is fixed to the
		fiducial cosmology $\{\Omega_{m0}=0.3098, \sigma_8=0.75, w=-1.0 \}$, while
		the remaining cases assume covariance matrices fixed to shifted
		cosmological models, as indicated in the legend.}
	\label{fig:MCMC_redshift_comparison}
\end{figure}

\subsection{$S_8$ tension}
\label{sec:s8_tension}

Building on the preceding sections, where the parameter estimators are unbiased
(section~\ref{sec:mc}) while the posterior widths are mis-estimated by tens of per cent
(section~\ref{sec:s_matrix}), we now examine the consequences for the $S_8$ tension
between cluster counts and early-time probes.

Because covariance misspecification leaves the recovered parameters in place and acts
only on their uncertainties, it cannot create or move a tension; it rescales the
tension's \emph{statistical significance}. The relevant axis is $S_8 \equiv
	\sigma_8\sqrt{\Omega_m/0.3}$, which lies approximately orthogonal to the principal
$(\Omega_m,\sigma_8)$ degeneracy of cluster observables and is therefore the
best-constrained combination in the cluster posterior. It is the natural parameter
along which the consistency between cluster counts and CMB constraints is assessed, and
it is the $S_8$ error that directly sets the significance of the tension. Concretely,
the tension significance between clusters (A) and the CMB (B) is
\begin{equation}
	n_\sigma = \frac{|S_8^{(A)} - S_8^{(B)}|}{\sqrt{\sigma_A^2 + \sigma_B^2}},
	\label{eq:nsigma}
\end{equation}
\citep{Pantos2026}. When the cluster uncertainty dominates ($\sigma_A \gg \sigma_B$),
a fractional mis-estimate $\sigma_A \to f\,\sigma_A$ maps any true tension $n_\sigma$
to a reported value $n_\sigma^{\rm rep} = n_\sigma / f$: a factor-of-$f$ error in
$\sigma_A$ produces an identical factor in the stated significance.

From table~\ref{tab:s8-std-setup}, the largest distortion of the marginal $S_8$
standard deviation occurs under \textsc{full-fix} with $\Delta\sigma_8 = +0.10$, where
the $S_8$ error is overestimated by $+32\%$, and with $\Delta\sigma_8 = -0.10$, where
it is underestimated by $-33\%$ (idealised, no-proxy case). The marginal $\sigma_8$
uncertainty is affected even more strongly in the same cases ($+41\%$ and $-39\%$). However,
as discussed above, $\sigma_8$ conflates the parameter-degeneracy direction with the
direction of the observed tension, making it a less relevant summary statistic. The
shot-noise-only (\textsc{no-SSC}) limit underestimates the $S_8$ error by $\sim44\%$.
A similar result holds when we include the mass-proxy and photo-z scatter: the \textsc{full-fix} distortion of the $S_8$
standard deviation reaches $+34\%/{-}30\%$ for the mass-proxy catalog (mp)
and $+36\%/{-}30\%$ for the mass-proxy-plus-photo-z catalog (mpz), both
essentially unchanged from the idealised case (table~\ref{tab:s8-std-setup}).

The sign follows the assumed amplitude: a covariance fixed at too low a fluctuation
amplitude underestimates $\sigma_A$, inflating the apparent tension, while too high an
amplitude softens it. Applying eq.~\eqref{eq:nsigma} in the cluster-dominated limit, a
genuine $2\sigma$ $S_8$ tension with $f=1.32$ ($+32\%$ overestimate) would be reported as
$2/1.32\approx1.5\sigma$, while $f=0.67$ ($-33\%$ underestimate) gives
$2/0.67\approx3.0\sigma$. Thus, the same underlying discrepancy can appear
substantially weaker or stronger solely because of the cosmology assumed when fixing
the covariance.

\begin{table}[ht]
	\centering
	\caption{Marginal standard deviations of $S_8$ and $\sigma_8$ at the representative
		survey layout ($\Omega_{\rm sky}=18000\,\mathrm{deg}^2$, $z_{\max}=0.8$), for
		each covariance specification (\textsc{S-fix} and \textsc{full-fix}), proxy
		class (no-proxy, mp, mpz), and the largest parameter
		displacements used in this analysis ($|\Delta\Omega_{m0}|=0.06$,
		$|\Delta\sigma_8|=0.10$, $|\Delta w|=0.40$). The ratio columns give the
		fractional change relative to the correctly-specified (\textsc{match},
		\textsc{S-fix}) run within each proxy class; that reference row is marked
		`ref'. Positive ratios indicate an overestimate of the uncertainty, negative
		ratios an underestimate. Note that the largest distortions of the $S_8$
		error are due to shifts in $\sigma_8$.}
	\label{tab:s8-std-setup}
	{\footnotesize
		\begin{tabular}{l l l r r r r r}
			\hline
			covariance        & proxy & configuration       & shift & $\mathrm{std}(S_8)$ & ratio   & $\mathrm{std}(\sigma_8)$ & ratio   \\
			\hline
			\textsc{S-fix}    & -     & \textsc{match}      & -     & 0.0072              & ref     & 0.0029                   & ref     \\
			\textsc{full-fix} & -     & \textsc{match}      & -     & 0.0072              & 0.3\%   & 0.0029                   & -0.3\%  \\
			\textsc{S-fix}    & -     & \textsc{no-SSC}     & -     & 0.0040              & -43.8\% & 0.0018                   & -36.4\% \\
			\textsc{full-fix} & -     & \textsc{no-SSC}     & -     & 0.0041              & -43.2\% & 0.0018                   & -37.3\% \\
			\textsc{S-fix}    & -     & $\Delta\Omega_{m0}$ & 0.06  & 0.0064              & -11.7\% & 0.0026                   & -10.5\% \\
			\textsc{full-fix} & -     & $\Delta\Omega_{m0}$ & 0.06  & 0.0080              & 11.8\%  & 0.0031                   & 8.2\%   \\
			\textsc{S-fix}    & -     & $\Delta\Omega_{m0}$ & -0.06 & 0.0088              & 21.8\%  & 0.0034                   & 18.0\%  \\
			\textsc{full-fix} & -     & $\Delta\Omega_{m0}$ & -0.06 & 0.0064              & -10.9\% & 0.0026                   & -9.8\%  \\
			\textsc{S-fix}    & -     & $\Delta\sigma_8$    & 0.10  & 0.0078              & 8.7\%   & 0.0031                   & 7.0\%   \\
			\textsc{full-fix} & -     & $\Delta\sigma_8$    & 0.10  & 0.0095              & 32.1\%  & 0.0041                   & 41.0\%  \\
			\textsc{S-fix}    & -     & $\Delta\sigma_8$    & -0.10 & 0.0066              & -8.7\%  & 0.0027                   & -6.7\%  \\
			\textsc{full-fix} & -     & $\Delta\sigma_8$    & -0.10 & 0.0048              & -32.7\% & 0.0018                   & -38.9\% \\
			\textsc{S-fix}    & -     & $\Delta w$          & 0.40  & 0.0079              & 10.1\%  & 0.0033                   & 13.5\%  \\
			\textsc{full-fix} & -     & $\Delta w$          & 0.40  & 0.0064              & -10.7\% & 0.0025                   & -12.7\% \\
			\textsc{S-fix}    & -     & $\Delta w$          & -0.40 & 0.0066              & -7.6\%  & 0.0027                   & -8.7\%  \\
			\textsc{full-fix} & -     & $\Delta w$          & -0.40 & 0.0079              & 9.4\%   & 0.0031                   & 6.9\%   \\
			\hline
			\textsc{S-fix}    & mp    & \textsc{match}      & -     & 0.0079              & ref     & 0.0034                   & ref     \\
			\textsc{full-fix} & mp    & \textsc{match}      & -     & 0.0081              & 2.8\%   & 0.0034                   & -0.1\%  \\
			\textsc{S-fix}    & mp    & \textsc{no-SSC}     & -     & 0.0064              & -18.6\% & 0.0032                   & -5.4\%  \\
			\textsc{full-fix} & mp    & \textsc{no-SSC}     & -     & 0.0065              & -17.1\% & 0.0033                   & -4.4\%  \\
			\textsc{S-fix}    & mp    & $\Delta\Omega_{m0}$ & 0.06  & 0.0074              & -5.6\%  & 0.0034                   & -0.8\%  \\
			\textsc{full-fix} & mp    & $\Delta\Omega_{m0}$ & 0.06  & 0.0094              & 18.9\%  & 0.0039                   & 15.2\%  \\
			\textsc{S-fix}    & mp    & $\Delta\Omega_{m0}$ & -0.06 & 0.0088              & 11.5\%  & 0.0035                   & 4.2\%   \\
			\textsc{full-fix} & mp    & $\Delta\Omega_{m0}$ & -0.06 & 0.0068              & -13.8\% & 0.0029                   & -14.7\% \\
			\textsc{S-fix}    & mp    & $\Delta\sigma_8$    & 0.10  & 0.0083              & 4.8\%   & 0.0035                   & 1.9\%   \\
			\textsc{full-fix} & mp    & $\Delta\sigma_8$    & 0.10  & 0.0105              & 33.9\%  & 0.0048                   & 39.7\%  \\
			\textsc{S-fix}    & mp    & $\Delta\sigma_8$    & -0.10 & 0.0076              & -3.5\%  & 0.0034                   & -0.4\%  \\
			\textsc{full-fix} & mp    & $\Delta\sigma_8$    & -0.10 & 0.0055              & -30.1\% & 0.0023                   & -32.1\% \\
			\textsc{S-fix}    & mp    & $\Delta w$          & 0.40  & 0.0081              & 2.8\%   & 0.0035                   & 3.7\%   \\
			\textsc{full-fix} & mp    & $\Delta w$          & 0.40  & 0.0072              & -8.7\%  & 0.0030                   & -11.2\% \\
			\textsc{S-fix}    & mp    & $\Delta w$          & -0.40 & 0.0077              & -2.5\%  & 0.0034                   & -0.3\%  \\
			\textsc{full-fix} & mp    & $\Delta w$          & -0.40 & 0.0088              & 11.1\%  & 0.0037                   & 8.5\%   \\
			\hline
			\textsc{S-fix}    & mpz   & \textsc{match}      & -     & 0.0083              & ref     & 0.0034                   & ref     \\
			\textsc{full-fix} & mpz   & \textsc{match}      & -     & 0.0087              & 4.0\%   & 0.0034                   & -0.2\%  \\
			\textsc{S-fix}    & mpz   & \textsc{no-SSC}     & -     & 0.0068              & -18.7\% & 0.0033                   & -4.5\%  \\
			\textsc{full-fix} & mpz   & \textsc{no-SSC}     & -     & 0.0070              & -16.2\% & 0.0032                   & -4.8\%  \\
			\textsc{S-fix}    & mpz   & $\Delta\Omega_{m0}$ & 0.06  & 0.0079              & -5.5\%  & 0.0034                   & -1.5\%  \\
			\textsc{full-fix} & mpz   & $\Delta\Omega_{m0}$ & 0.06  & 0.0100              & 20.1\%  & 0.0039                   & 15.5\%  \\
			\textsc{S-fix}    & mpz   & $\Delta\Omega_{m0}$ & -0.06 & 0.0094              & 13.0\%  & 0.0036                   & 4.1\%   \\
			\textsc{full-fix} & mpz   & $\Delta\Omega_{m0}$ & -0.06 & 0.0072              & -13.1\% & 0.0029                   & -16.1\% \\
			\textsc{S-fix}    & mpz   & $\Delta\sigma_8$    & 0.10  & 0.0089              & 6.4\%   & 0.0035                   & 1.8\%   \\
			\textsc{full-fix} & mpz   & $\Delta\sigma_8$    & 0.10  & 0.0114              & 36.4\%  & 0.0047                   & 37.5\%  \\
			\textsc{S-fix}    & mpz   & $\Delta\sigma_8$    & -0.10 & 0.0081              & -2.4\%  & 0.0034                   & -1.3\%  \\
			\textsc{full-fix} & mpz   & $\Delta\sigma_8$    & -0.10 & 0.0058              & -30.1\% & 0.0023                   & -32.9\% \\
			\textsc{S-fix}    & mpz   & $\Delta w$          & 0.40  & 0.0087              & 4.8\%   & 0.0035                   & 1.3\%   \\
			\textsc{full-fix} & mpz   & $\Delta w$          & 0.40  & 0.0075              & -9.4\%  & 0.0030                   & -12.0\% \\
			\textsc{S-fix}    & mpz   & $\Delta w$          & -0.40 & 0.0083              & -0.2\%  & 0.0034                   & -0.7\%  \\
			\textsc{full-fix} & mpz   & $\Delta w$          & -0.40 & 0.0094              & 12.9\%  & 0.0037                   & 8.0\%   \\
			\hline
		\end{tabular}
	}
\end{table}

\begin{table}[ht]
	\centering
	\caption{Relative difference in the credible-region volume, $\Delta V /
			V_{\text{match}}$, for different fixed-covariance cosmologies relative to
		the fiducial $(\Omega_{m0}, \sigma_8, w) = (0.3098, 0.75, -1)$. The triplet
		$(\Delta_{\Omega_m}, \Delta_{\sigma_8}, \Delta_w)$ denotes the fractional
		change in the marginal standard deviation of $(\Omega_{m0}, \sigma_8, w)$
		relative to the matched-covariance run (see eq.~\eqref{eq:width_ratio}).
		The upper group keeps $\Omega_{m0}$ at its fiducial value while varying
		$\sigma_8$ (and hence $S_8$); the lower group holds $S_8 = 0.762$ fixed and
		moves along the $(\Omega_{m0}, \sigma_8)$ degeneracy.}
	\label{tab:parameter_comparison}
	{
		\begin{tabular}{|c|c|c|c|}
			\hline
			$(\Omega_{m0}, \sigma_8)$ & $(\Delta_{\Omega_m}, \Delta_{\sigma_8}, \Delta_w)$ & $\Delta V / V_{\text{match}}$ (true) & $\Delta V / V_{\text{match}}$ (mpz) \\ \hline\hline
			\multicolumn{4}{|c|}{$\Omega_{m0} = 0.3098$ (fiducial); $\sigma_8$ varied ($S_8$ changes)}                                                                  \\ \hline
			$(0.3098, 0.60)$          & $(-61\%, -60\%, -66\%)$                            & $-96\%$                              & $-88\%$                             \\ \hline
			$(0.3098, 0.70)$          & $(-22\%, -22\%, -25\%)$                            & $-58\%$                              & $-46\%$                             \\ \hline
			$(0.3098, 0.80)$          & $(+22\%, +22\%, +26\%)$                            & $+101\%$                             & $+75\%$                             \\ \hline
			$(0.3098, 0.85)$          & $(+45\%, +43\%, +56\%)$                            & $+272\%$                             & $+178\%$                            \\ \hline\hline
			\multicolumn{4}{|c|}{$S_8 = 0.762$ fixed; $(\Omega_{m0}, \sigma_8)$ varied along the degeneracy}                                                            \\ \hline
			$(0.4850, 0.60)$          & $(-32\%, -31\%, -41\%)$                            & $-81\%$                              & $-51\%$                             \\ \hline
			$(0.3567, 0.70)$          & $(-11\%, -11\%, -15\%)$                            & $-40\%$                              & $-24\%$                             \\ \hline
			$(0.2734, 0.80)$          & $(+14\%, +13\%, +19\%)$                            & $+67\%$                              & $+31\%$                             \\ \hline
			$(0.2423, 0.85)$          & $(+30\%, +29\%, +41\%)$                            & $+173\%$                             & $+76\%$                             \\ \hline
		\end{tabular}
	}
\end{table}

Up to this point we have considered single-parameter sweeps and found that variations
in $\sigma_8$ produce the largest covariance-induced distortions. However, changing
$\sigma_8$ also changes $S_8 \equiv \sigma_8\sqrt{\Omega_m/0.3}$, the parameter
combination that is most tightly constrained by cluster counts and along which tensions
with early-time probes are commonly quantified. To disentangle these effects, we extend
the analysis in two directions. First, we increase the range of the $\sigma_8$ scan
from the $\Delta\sigma_8=\pm0.10$ variations considered above to
$\sigma_8=0.60$--$0.85$ (table~\ref{tab:parameter_comparison}, upper), establishing a baseline
for the large-mismatch regime. Second, we repeat the scan while adjusting $\Omega_{m0}$ so
as to keep $S_8=0.762$, equal to its fiducial value (table~\ref{tab:parameter_comparison}, lower). This
comparison isolates the extent to which the covariance response is associated with
changes in $S_8$ itself, rather than with variations in $\sigma_8$ at fixed $\Omega_{m0}$.
Moreover, we focus on an LSST-like catalog with $z_{\rm max}=1.5$ and an area of
$18\,000\,\mathrm{deg}^2$, and consider two observational scenarios: one using true
masses and redshifts, and another incorporating mass-proxy and photometric-redshift
uncertainties.

To probe the large-mismatch regime, we replace the marginal standard deviation with the
full credible-region volume $\Delta V/V_{\rm match}$ and scan fixed covariance
amplitudes over the wide range $\sigma_8 = 0.60$--$0.85$, bracketing the true fiducial
$(\Omega_{m0}, \sigma_8, w) = (0.3098, 0.75, -1)$, for both idealised and
mass-proxy-plus-photo-z catalogs (figure~\ref{fig:MCMC_redshift_comparison}). We
perform this scan in two complementary ways (18 MCMC analyses): (i) varying $\sigma_8$ alone, so that
$S_8$ also changes (table~\ref{tab:parameter_comparison}, upper), and (ii) moving along
the $(\Omega_{m0}, \sigma_8)$ degeneracy at fixed $S_8 = 0.762$
(table~\ref{tab:parameter_comparison}, lower), which isolates the $S_8$-aligned
response identified in section~\ref{sec:s_matrix}.

The credible volume now changes by $-88\%$ to $+178\%$ across the assumed amplitudes in
the mpz case (table~\ref{tab:parameter_comparison}, upper), and for the most
discrepant covariance ($\sigma_8=0.60$) the posterior is compressed so strongly
that, for this realization, the true cosmology falls outside the $95\%$ region for
both no-proxy and mpz cases (see left panels of
figure~\ref{fig:MCMC_redshift_comparison}). The offset of
the posterior maximum with respect to the true cosmology is set by the particular
realization, as established in section~\ref{sec:mc}; what the covariance choice controls
is the width of the credible region, and shrinking it is enough to turn a statistically
unremarkable offset into an apparent tension, with no change in the underlying
model. The comparison with the fixed-$S_8$ construction confirms that the effect is
governed by the mismatch in $S_8$: holding $S_8$ fixed while moving along the
$(\Omega_{m0},\sigma_8)$ degeneracy roughly halves the volume change (for example
$-88\%\to-51\%$ at the lowest amplitude in the mpz case;
table~\ref{tab:parameter_comparison}, lower), consistent with the $S_8$-aligned
response identified in section~\ref{sec:s_matrix}, see also the right panels of
figure~\ref{fig:MCMC_redshift_comparison}. Including mass-proxy and
photometric-redshift uncertainties further reduces the discrepancy, in line with
section~\ref{sec:realism}.

These results show that neglecting the cosmology dependence of the covariance can materially
misstate the significance of the $S_8$ tension between early- and late-time probes,
even though it leaves the recovered cosmology unchanged. A consistent cosmology-dependent
treatment of the covariance is therefore essential not only for parameter inference but also
for robust assessments of cosmological tensions in the precision-cosmology era.

\subsection{Covariance matrix fixed to the best-fit}
\label{sec:best_fit}

Given that the dominant impact of fixing the covariance matrix concerns the inferred
parameter uncertainties rather than the location of the maximum likelihood itself, we
now explore an iterative strategy in which the covariance matrix is recalculated at the
recovered best-fit cosmology. The motivation for this approach is that the previous
analyses indicate that covariance misspecification primarily distorts the uncertainty
normalization and the geometry of the credible regions, while leaving the
cosmological parameter estimators nearly unbiased.

Specifically, we first estimate the best-fit parameters using a fixed covariance
matrix, and subsequently recompute the covariance matrix at the recovered best-fit
values in order to derive the final cosmological constraints.

For this test, we consider the mirror-analysis configuration discussed in the previous sections. We generate mock realizations at the cosmologies $(\Omega_{m0}, \sigma_8, w) = (0.3098, 0.75, -1.0)$ and $(0.6486, 0.95, -2.0)$, while fixing the covariance matrix to that of the fiducial $\Lambda$CDM model of eq.~\eqref{eq:fiducial_main}. The first cosmology corresponds to the fiducial model adopted in section~\ref{sec:s8_tension} and differs from the reference cosmology only through the value of $\sigma_8$, thereby providing a direct connection with the results presented there. The second cosmology represents a more extreme departure from the fiducial model and probes a different region
of parameter space. For each cosmology, we generate a single realization assuming a
LSST-like survey configuration with area $\Omega_{\mathrm{sky}} = 18000\,\mathrm{deg}^2$
and maximum redshift $z_{\max}=1.5$. The analyses are performed for both idealized
(no-proxy) and more realistic (mpz) catalogs with $\sigma^{\mathrm{ph}}_0
	= 0.03$.

\begin{figure}[ht]
	\centering

	\includegraphics[width=0.48\linewidth]{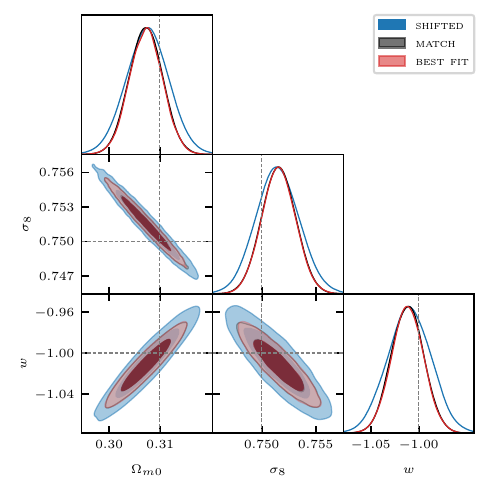}
	\includegraphics[width=0.48\linewidth]{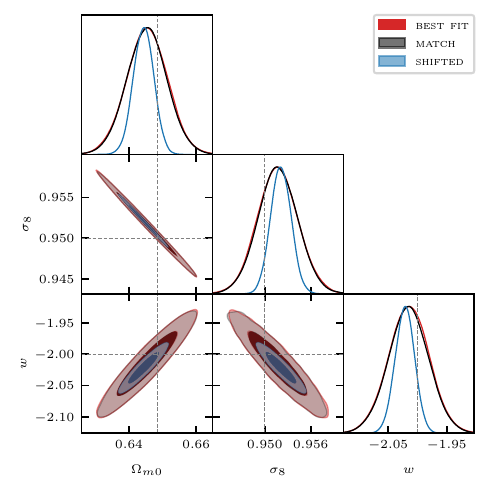}\\
	\includegraphics[width=0.48\linewidth]{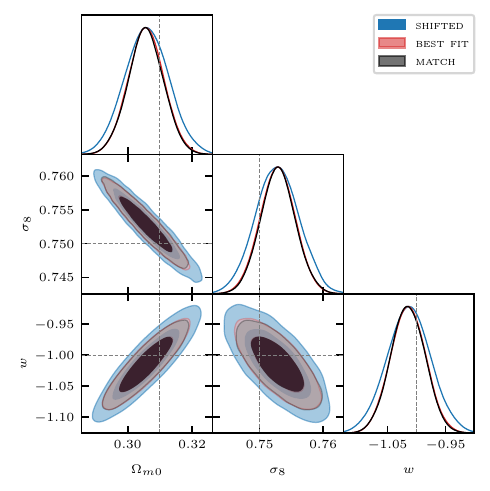}
	\includegraphics[width=0.48\linewidth]{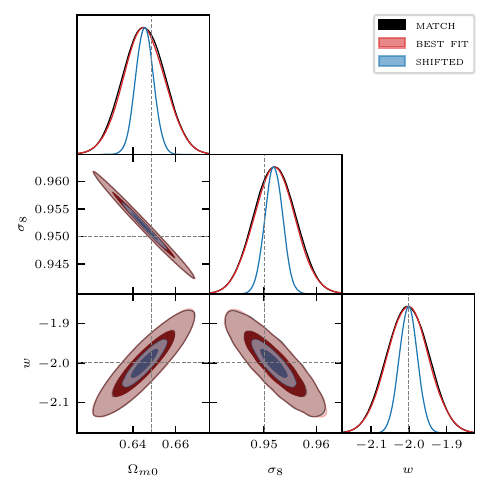}

	\caption{Corner plots for the cosmological parameters $\{\Omega_{m0}, \sigma_8,
			w\}$ comparing analyses performed with a matched covariance (gray), a
		shifted covariance fixed at the fiducial model of
		eq.~\eqref{eq:fiducial_main} (blue), and an iterated covariance obtained by
		recomputing the covariance at the recovered best-fit cosmology (red). The
		top(bottom) panels show the no-proxy(mpz) analyses. The left and
		right panels show the mirror-analysis realizations generated at
		$(\Omega_{m0}, \sigma_8, w) = (0.3098, 0.75, -1.0)$ and $(0.6486, 0.95,
			-2.0)$, respectively, following table~\ref{tab:cov_iter}. Contours denote
		the 68\% and 95\% credible regions. }
	\label{fig:MCMC_bf_comparison}
\end{figure}

The results, summarized in table~\ref{tab:cov_iter}, demonstrate that the relative
difference in credible-region volume, $\Delta V/V_{\mathrm{match}}$, is dramatically
reduced once the covariance matrix is recomputed at the recovered best-fit cosmology.
For the iterated covariance, table~\ref{tab:cov_iter} reports the volume differences of
both the $1\sigma$ and $2\sigma$ credible regions, see also
figure~\ref{fig:MCMC_bf_comparison}. Unlike most of the models considered in this work,
these cases exhibit a small but noticeable dependence on the posterior probability level,
indicating that the covariance mismatch affects the tails of the posterior somewhat
differently from the central region. The sign of the discrepancy is not universal: for the mild displacement, the misspecified covariance \emph{inflates} the credible-region volume by $151\%$ (no-proxy) and $106\%$ (mpz), whereas for the extreme displacement, it reduces the volume by $90\%$ and $92\%$, respectively. A misspecified covariance is therefore not reliably conservative; whether it over- or underestimates the credible-region volume depends on how far the assumed cosmology
sits from the truth. Nevertheless, in every case a single covariance update reduces the
volume discrepancy to a few percent, at most $+5\%$ at $2\sigma$. Since the
credible-region volume depends on the full parameter covariance matrix, including all
variances and correlations, the corresponding changes in the marginalized uncertainties
of individual parameters are expected to be even smaller.

This confirms that a single covariance update at the recovered best-fit cosmology is
sufficient to restore the correct uncertainty normalization. The reason is that the
cosmology dependence of the covariance is smooth on the scale of the parameter
uncertainties: from the slope of the \textsc{full-fix} response in
section~\ref{sec:s_matrix} (a $+41\%$/$-39\%$ change in the $\sigma_8$ width over
$\Delta\sigma_8=\pm0.10$), a $1\sigma$ error in the recovered $\sigma_8$ shifts the
inferred widths by about $1\%$, and only a few percent even at $3\sigma$. A single
update therefore lands the covariance close enough to the truth that the residual
distortion is negligible. In practice, fixing the covariance matrix at a statistically
supported cosmology provides an effective approximation, reducing computational cost
while maintaining accurate parameter uncertainties and improving MCMC convergence.

\begin{table}[ht]
	\centering
	\caption{Relative difference in credible-region volume, $\Delta
			V/V_{\mathrm{match}}$, for analyses performed with a shifted covariance and
		after one covariance iteration. The iterated covariance is obtained by
		recomputing the covariance at the recovered best-fit parameters and
		rerunning the analysis. Before the iteration, the covariance is fixed at
		the fiducial model given in eq.~\eqref{eq:fiducial_main}.}
	{\small\setlength{\tabcolsep}{4pt}
	\begin{tabular}{lccccc}
		\hline
		Model                &
		Simulation cosmology &
		Recovered best fit   &
		Shifted              &
		Iterated ($1\sigma$) &
		Iterated ($2\sigma$)                                                                                   \\
		\hline
		\multirow{2}{*}{no-proxy}
		                     & $(0.3098, 0.75, -1.0)$ & $(0.3075, 0.751, -1.010)$ & $+151\%$ & $+1\%$ & $+4\%$ \\
		                     & $(0.6486, 0.95, -2.0)$ & $(0.6444, 0.952, -2.022)$ & $-90\%$  & $+2\%$ & $+3\%$ \\
		\hline
		\multirow{2}{*}{mpz}
		                     & $(0.3098, 0.75, -1.0)$ & $(0.3057, 0.753, -1.014)$ & $+106\%$ & $+2\%$ & $+5\%$ \\
		                     & $(0.6486, 0.95, -2.0)$ & $(0.6455, 0.952, -2.001)$ & $-92\%$  & $-1\%$ & $-2\%$ \\
		\hline
	\end{tabular}
	}
	\label{tab:cov_iter}
\end{table}

Two caveats qualify this conclusion. First, the procedure recomputes the covariance at
the best fit, so it relies on that best fit being close to the truth, which in turn
relies on the data constraining the cosmology tightly. The forecasts here neglect
completeness, purity and other observational systematics, so the constraints we report
are optimistic; with a fuller systematic budget the cluster posterior broadens, the
best fit can fluctuate further from the truth, and the recomputed covariance becomes a
correspondingly poorer estimate. The smoothness of $C(\theta)$ keeps even a
several-$\sigma$ excursion of the best-fit at the few-percent level, but the
sub-percent margin quoted above is specific to our error budget and should be
re-validated against the actual, systematics-inflated constraint of a given survey.
Because $C(\hat\theta)$ is a smooth model function of the best fit rather than a
covariance estimated from the data, it requires no Hartlap-type correction, and the
residual data dependence, entering only through $\hat\theta$, is second order and
negligible.

Second, in a multi-probe analysis, the cluster covariance is more consistently  recomputed at the \emph{cluster-counts-only} best fit than  at a joint one, since a combined fit would place $C$ at
a cosmology that the clusters alone do not prefer. This consideration holds even in the absence of any tension. Where early- and late-time probes are genuinely discrepant, the single best fit  becomes ambiguous, and a fully cosmology-dependent covariance, capable of consistently accommodating the parameter differences, remains necessary
(section~\ref{sec:s8_tension}).

\section{Conclusions}
\label{sec:conclusions}

In this work we have investigated how assumptions about the covariance matrix of galaxy
cluster number counts propagate into the inference of cosmological parameters, in
particular $\Omega_{m0}$, $\sigma_8$, and $w$. Cluster abundances provide strong
constraints on cosmology, but their likelihood depends not only on the mean counts but
also on their covariance, which includes contributions from Poisson shot noise and
super-sample covariance (SSC) induced by long-wavelength matter density fluctuations.
Since both the signal and its covariance depend on cosmology, parameter inference
ideally requires a fully cosmology-dependent treatment of the likelihood. In practice,
however, the covariance, and especially its expensive SSC term, is often evaluated at a
single fixed fiducial model for reasons of computational cost.

To quantify the consequences, we re-analysed a single realistic fiducial mock per survey
layout under a controlled set of covariance constructions: fixing only the SSC kernel
$S$ while recomputing the abundance and bias prefactors (\textsc{S-fix}), fixing the
entire covariance (\textsc{full-fix}), and dropping the SSC term altogether
(\textsc{no-SSC}), each evaluated either at the true cosmology (\textsc{match}) or at a
displaced one. Because all variants act on the \emph{same} data vector, any change is
attributable solely to the covariance choice. We probed the effect in two equivalent
ways, a \emph{forward} configuration that displaces the covariance on a fiducial mock and
a \emph{mirror} configuration that displaces the data while holding the
covariance at the fiducial value, and we characterised it through individual MCMC
posterior widths and a Monte Carlo ensemble of best fits.

The cosmological parameter estimators remain unbiased under covariance misspecification.
Across the entire scan over displacements, layouts and covariance constructions, the
recovered means depart from the reference analysis by at most $2.2\%$ for $w$,
$0.5\%$ for $\Omega_{m0}$ and $0.25\%$ for $\sigma_8$ with the default binning, never
exceeding $0.3\,\sigma$ of the parameter uncertainty, and by at most $0.6\,\sigma$ in
the finer-binned repeats; a Monte Carlo ensemble of $N=1000$ mocks confirms a residual
bias below $0.15\,\sigma$ in the correctly specified and misspecified
configurations. The dominant impact of covariance misspecification is therefore not a
shift in the best-fit parameters, but a modification of the inferred parameter
uncertainties and of the geometry of the credible regions.

Fixing the covariance is not, in itself, the source of error: under \textsc{full-fix}
at the \emph{true} cosmology, the posterior widths track \textsc{match} to within
$2.4\%$ ($R_\theta\simeq1$ for every parameter and survey layout). The error arises
only when the covariance is fixed at the \emph{wrong} cosmology, and its size depends strongly on which ingredients are frozen. Freezing only the $S$ matrix (\textsc{S-fix}) while keeping the counts
and bias consistent has a moderate impact, with width changes $\Delta_\theta \lesssim7\%$ for $\sigma_8$ displacements and up to $+23\%$ for the largest $\Omega_{m0}$
displacements. Fixing the full covariance (\textsc{full-fix}), by contrast, distorts the widths substantially and monotonically: for $\sigma_8$ we find a $+41\%$/$-39\%$ mis-estimation at
$\Delta\sigma_8=\pm0.10$, a covariance built at a lower (higher) power spectrum
amplitude yielding artificially tight (broad) posteriors. Since the $S$ matrix is the
computationally expensive ingredient, the robustness of \textsc{S-fix} is a useful
practical result.

The contrast between \textsc{S-fix} and \textsc{full-fix} identifies where the cosmology
dependence of the covariance actually resides. Their only difference is whether the
abundance and bias prefactors $b\,\bar\mu$ are recomputed or frozen, so the fact that
\textsc{S-fix} distorts the widths considerably less than \textsc{full-fix} shows that the
dominant cosmology dependence of the cluster-count covariance lies in these prefactors rather than
in the $S$ matrix itself. This is physically expected: the cluster abundance is
exponentially sensitive to the amplitude of matter fluctuations, so freezing $b\,\bar\mu$
at a wrong amplitude strongly mis-scales the SSC term, whereas $S$ varies comparatively
slowly with cosmology.

Although variations in $\Omega_{m0}$ and $\sigma_8$ do not produce a simple monotonic
behaviour when considered individually, the dominant trends are organised by the derived
parameter $S_8$. Since cluster counts primarily constrain a degenerate combination of
$\Omega_{m0}$ and $\sigma_8$, the effect of covariance misspecification projects mainly
along this direction and acts as an effective rescaling of the uncertainties along the
$S_8$ degeneracy axis. Variations in $w$ have a subdominant effect. The survey
area, by contrast, is not decisive: because $S_{\alpha\beta}\propto1/f_{\rm sky}$, the
super-sample and shot-noise terms carry the same leading area scaling and their relative
weight is nearly independent of the footprint. Consistently, the \textsc{no-SSC}
construction underestimates the $\sigma_8$ uncertainty by $36\%$ at
$\Omega_{\rm sky}=18000\,\mathrm{deg}^2$ and by $40\%$ at $3000\,\mathrm{deg}^2$,
i.e.\ by $35$--$40\%$ in every layout considered, with only a weak dependence on
depth.

The \emph{forward} and \emph{mirror} configurations yield the same result once both are
expressed in terms of the mismatch between the data and covariance cosmologies. The
impact of covariance misspecification depends only on this mismatch and not on which of
the two is displaced, so the controlled forward scan is representative of a real analysis,
in which the covariance is fixed at an assumed cosmology while the data correspond to the
unknown true one. As a complementary measure to the marginal widths we also report the
joint credible volume, which additionally captures changes in the correlation structure
of the parameters: under \textsc{full-fix} it changes by $+116\%$ and $-72\%$ at
$\Delta\sigma_8=\pm0.10$, and the same volume is recovered either as the highest-posterior
-density volume of the MCMC chains or as $\sqrt{\det\mathrm{Cov}}$ of the Monte Carlo
best-fit ensemble.

These results are robust to observational realism. Repeating the analysis with a
mass proxy and with LSST-like photometric redshifts broadens the constraints by up to
$\sim30\%$ but leaves the covariance-misspecification effects essentially unchanged: the
\textsc{full-fix} distortion stays at the $+38\%$/$-33\%$ level on $\sigma_8$ and the estimators
remain unbiased, while the shot-noise-only (\textsc{no-SSC}) underestimation is the one effect
that mildly relaxes, as the proxy scatter lowers the relative weight of the SSC term.

This width effect propagates directly into cosmological tension. Because covariance
misspecification rescales the parameter uncertainties without moving the estimators, it
rescales the statistical \emph{significance} of the $\sigma_8$/$S_8$ tension between cluster
counts and early-time probes without creating or removing it. In the regime where the
cluster measurement dominates the combined uncertainty, the $+32\%$/$-33\%$ mis-estimation of the
$S_8$ width turns a nominal $2\sigma$ tension into anywhere from $\sim1.5\sigma$ to
$\sim3.0\sigma$ depending only on the cosmology assumed for the covariance, and in extreme
cases the distortion is large enough to place the true cosmology outside the $95\%$ region.

A simple mitigation is available for single-probe analyses: because the estimators are
unbiased and the covariance varies smoothly on the scale of the parameter uncertainties (a
$1\sigma$ error in $\sigma_8$ changes the widths by about $1\%$), recomputing the
covariance once at the recovered best fit restores the correct uncertainties even when
starting from a badly misspecified covariance. Two qualifications apply. The constraints
forecast here are optimistic, as they omit cluster completeness and purity and other
observational systematics; a wider, more realistic posterior lets the best fit fluctuate
further from the truth, so the sub-percent margin should be re-validated against a survey's
actual constraint. And in a multi-probe setting the cluster covariance is more
consistently evaluated at the cluster-counts-only best fit than at a joint one, since a
combined fit would place $C$ at a cosmology that the clusters alone do not prefer; where
the probes are genuinely discrepant, no single best fit serves them all and a fully
cosmology-dependent covariance remains the appropriate treatment.

For multi-probe analyses combining early- and late-time observables, a fully
cosmology-dependent treatment of the covariance matrix remains necessary to ensure
internal consistency and to avoid propagating mismatches between probes. Overall, our
results demonstrate that the dominant cosmology dependence of the cluster-count covariance
arises from the abundance and bias prefactors and their sensitivity to the amplitude of
matter fluctuations, that this dependence projects naturally onto the $S_8$ degeneracy
direction, and that a consistent cosmology-dependent treatment of the covariance matrix is
essential for robust cluster cosmology analyses and for reliable assessments of
concordance or tension between early- and late-time probes in the precision era of
large-scale structure surveys.


\appendix

\section{Estimation of Posterior Credible Volumes from MCMC Chains}
\label{app:HDP}

In this work, parameter constraints are quantified through the volume of the
$N$-dimensional highest-posterior-density (HPD) region enclosing a credible fraction
$\alpha$ of the posterior distribution. This appendix summarizes the numerical
procedure used to estimate this volume from MCMC samples.

Let $P(\vec{\theta}|D)$ denote the posterior distribution for the parameter vector
$\vec{\theta}$. In particular, $\vec{\theta} = \{\Omega_{m0}, \sigma_8, w\}$. The HPD region $\Omega_\alpha$ is defined such that

\begin{equation}
	\alpha =
	\frac{\int_{\Omega_\alpha} P(\vec{\theta}|D)\, d\Omega_\theta}
	{\int_{\Omega_{\mathrm{total}}} P(\vec{\theta}|D)\, d\Omega_\theta},
\end{equation}
where $\alpha$ typically takes the values $0.6827$ or $0.9545$. The associated credible
volume is

\begin{equation}
	V = \int_{\Omega_\alpha} d\Omega_\theta.
\end{equation}

In practice, the HPD region is approximated directly from the MCMC chain of size
$N_{\mathrm{chain}}$. The samples are sorted in decreasing order of posterior
probability (equivalently, increasing order of $-2\ln P$). The first
$$
	N = \mathrm{int}(\alpha N_{\mathrm{chain}})
$$
samples define the discrete approximation to $\Omega_\alpha$.

From this subset, we compute

\begin{equation}
	s = \sum_{i=1}^{N}
	\exp\left[
		\frac{-2\ln P_i - (-2\ln P_{\mathrm{bf}})}{2}
		\right],
\end{equation}
where $P_{\mathrm{bf}}$ denotes the posterior evaluated at the maximum (best-fit) point.

The logarithmic volume is then estimated as

\begin{equation}
	\label{eq:lnV_estimator}
	\ln V =
	\ln(P_{\mathrm{norm}})
	- \ln P_{\mathrm{bf}}
	+ \ln\left(\frac{s}{N_{\mathrm{chain}}}\right),
\end{equation}
where $P_{\mathrm{norm}}$ is the posterior normalization constant. The credible volume
$V$ is obtained by exponentiation.

This estimator provides a numerical approximation to the HPD credible volume directly
from the MCMC samples, avoiding assumptions about Gaussianity or parameter degeneracy
structure. It is implemented in \texttt{NumCosmo}~\cite{DiasPintoVitenti2014} as the
\texttt{get\_post\_lnvol} method of the Monte Carlo catalogue class, which we use to
compute all credible volumes reported in this work.

\subsection*{Gaussian limit and the Monte Carlo estimate}

The estimator of eq.~\eqref{eq:lnV_estimator} is non-parametric, but it is instructive to
connect it to the Gaussian approximation and to the Monte Carlo ensemble of
section~\ref{sec:mc}. If the posterior is approximately Gaussian with covariance $C$ in
$d$ dimensions, the HPD region at level $\alpha$ is the ellipsoid
$\{\vec\theta:(\vec\theta-\vec\mu)^\top C^{-1}(\vec\theta-\vec\mu)\le\chi^2_d(\alpha)\}$,
with $\chi^2_d(\alpha)$ the corresponding quantile of the chi-squared distribution with
$d$ degrees of freedom. Its volume is
\begin{equation}
	\label{eq:gaussian_volume}
	V_\alpha = \frac{\pi^{d/2}}{\Gamma\!\left(d/2+1\right)}
	\left[\chi^2_d(\alpha)\right]^{d/2}\sqrt{\det C},
\end{equation}
so that, at fixed credible level and dimension, $V_\alpha\propto\sqrt{\det C}$.

Equation~\eqref{eq:gaussian_volume} provides a second, independent route to the same
volume. From the Monte Carlo ensemble of section~\ref{sec:mc} we form the sample
covariance $\hat C$ of the $N$ best-fit estimates and evaluate $\sqrt{\det\hat C}$. Because
the chain-based estimator of eq.~\eqref{eq:lnV_estimator} makes no Gaussianity assumption,
whereas $\sqrt{\det\hat C}$ does through eq.~\eqref{eq:gaussian_volume}, their agreement
(section~\ref{sec:mc}) simultaneously confirms that the posterior is close to Gaussian over
the relevant region and that the width of the MCMC posterior faithfully tracks the
frequentist scatter of the estimator across realizations. We therefore use the two readings
of the credible volume, from the posterior chain and from the ensemble of estimators,
interchangeably in the main text.







\acknowledgments{

	HCNL acknowledges the support of the Coordenação de Aperfeiçoamento de Pessoal de
	Nível Superior (CAPES) under grant 88887704526/2022-00. All authors acknowledge the
	support of Fundação Araucária (NAPI de Fenômenos Extremos do Universo, Grant No.\
	347/2024 PDI).}

\bibliographystyle{JHEP}
\bibliography{references_cov}

\end{document}